\def\bq{\begin{quote}}
\def\eq{\end{quote}}
\def \lsim{\mathrel{\vcenter
     {\hbox{$<$}\nointerlineskip\hbox{$\sim$}}}}

\def\gappeq{\mathrel{\rlap {\raise.5ex\hbox{$>$}}
{\lower.5ex\hbox{$\sim$}}}}
\def\lappeq{\mathrel{\rlap{\raise.5ex\hbox{$<$}}
{\lower.5ex\hbox{$\sim$}}}}
\documentstyle[axodraw]{frank}

\newenvironment{comment}[1]{}{}
\def\beq{\begin{equation}}
\def\eeq{\end{equation}}
\def\bea{\begin{eqnarray}}
\def\eea{\end{eqnarray}}
\def\barr{\begin{array}}
\def\earr{\end{array}}
\newcommand{\sm}{standard model}
\newcommand{\cm}{center of mass}
\newcommand{\xs}{cross section}
\newcommand{\lc}{linear collider}
\newcommand{\sw}{\mbox{$\sin\theta_w$}}
\newcommand{\cw}{\mbox{$\cos\theta_w$}}
\newcommand{\swt}{\mbox{$\sin^2\theta_w$}}

\newcommand{\pe}{\mbox{$e^+e^-$}}
\newcommand{\ee}{\mbox{$e^-e^-$}}
\newcommand{\ep}{\mbox{$e^-\gamma$}}
\newcommand{\pp}{\mbox{$\gamma\gamma$}}
\newcommand{\dil}{bilepton}
\newcommand{\ml}{maximum likelihood}

\begin{document}

\begin{flushright}
PSI-PR-96-21\\
MPI-PhT/96-45\\
September 1996
\end{flushright}

\vskip2cm

\begin{frontmatter}
\title{Bileptons: \\
        Present Limits and Future Prospects}
\author{Frank Cuypers}
\address{{\tt cuypers@pss058.psi.ch}\\
        Paul Scherrer Institute,
        CH-5232 Villigen PSI,
        Switzerland}
\author{Sacha Davidson}
\address{{\tt sacha@mppmu.mpg.de}\\
        Max-Planck-Institut f\"ur Physik,
        F\"ohringer Ring 6, 
        D-80805 M\"unchen,
        Germany}
\begin{abstract}
We define bileptons to be bosons coupling to a pair of leptons
and construct the most general dimension four lagrangian
involving scalar and vector bileptons.
We concentrate on fields with lepton number 2,
and  derive model independent bounds 
on their masses and couplings
from low-energy data.
In addition,
we study their signals 
in high energy experiments
and forecast the discovery potential of future colliders.
\end{abstract}

\begin{keyword}
	bileptons,
	dileptons,
	new physics,
	new bosons,
	low energy experiments,
	colliders experiments
\PACS
	14.80.-j, 
	12.60.-i, 
	12.10.Dm, 
	13.10.+q
\end{keyword}

\end{frontmatter}

\clearpage

\setcounter{tocdepth}{5}
\tableofcontents
\clearpage

\section{Introduction}

The \sm\ of strong and electroweak interactions 
describes present data very successfully. However, it is
commonly believed that it is not the end of the story: 
grand unified theories appeal to our craving for elegance, and 
naturalness arguments lead many theorists to believe
that there should be some new physics lurking at the
TeV scale. However, opinions differ on how to
extend the \sm.

One of the peculiar features of the  \sm\ is 
that none of its bosons 
carry global quantum numbers; only
the fermions carry baryon or lepton number.
This is no longer the case in
most  popular extensions.
For instance, the scalar spartners in the supersymmetric
\sm\ carry the same  baryon or lepton number
as their associated fermions. Grand unified theories,
technicolour, and compositeness scenarios often 
predict the existence of  light leptoquarks, di-  or \dil s,
and diquarks. These scalar or vector
bosons  respectively have baryon and/or 
lepton number conserving couplings   to a lepton and a quark,
two leptons, or two quarks.  

We define a \dil\ to be a boson
which couples minimally ({\it i.e.}, with dimension
four interactions) to \sm\ leptons
(we do not include right-handed
neutrinos), but not to quarks. 
These particles can carry lepton number 0 or 2.
They have previously shared the name ``dilepton''
with events having two final state leptons; to
reduce confusion, we call these particles
``bileptons'', following Frampton \cite{Frampton}.
We shall concentrate here on the $L=2$ \dil s,
since the $L=0$ \dil s have very similar properties 
as the familiar \sm\ bosons.
Our aim is to list the present bounds on
\dil s that can be calculated from
low-energy physics  and from LEP, and to
explore   bounds on \dil s from future
high energy experiments.
We  mainly deal with discovery limits 
and will not dwell on the issue
of uniquely determining the quantum numbers of the \dil s.

Bileptons are present in many extensions
of the \sm.  Scalars appear in models that
generate neutrino majorana masses (see,
for instance, \cite{CL,GR,Z,Pec}), and in
various theories with enlarged Higgs sectors
(such as left-right models \cite{R,GM,LR}). 
Massive gauge \dil s may appear 
 when the  \sm\
is embedded in a larger gauge group \cite{FL,Pal,F,F'},
and non-gauge vectors can appear in composite
and technicolour theories \cite{compTC}.

Various authors have
previously studied constraints
on \dil s. The low energy bounds were
computed  in a model independent
way in  \cite{DH,MRS,MSch}. Constraints  and possible collider 
signals for
specific models have been calculated in
\cite{BBKP,S,STU,IKMMY,MV,11,LTV,V,CK,CF,FN,FNS,FMSS,HS,H,HSS,GMS,GGMKO,HM,CMS,S',LTNL,opal,AP,FNKY}.  
 More
recently, the constraints on new physics following
from the improved $\tau$ data have been calculated
\cite{PS}.
In this paper, 
we catalogue the low- and high-energy constraints in as complete and
model independent a way as possible. We also present
the bounds assuming three representative models for
the generation structure of the \dil-fermion-fermion
coupling.

This work is divided into three parts.
First,
we write down a rather general class of \dil\ lagrangians,
which encompasses our definition of
{\em bosons which couple minimally to leptons but not to quarks}.
We then  derive the present low-energy bounds
on the \dil\ couplings and masses.
Finally,
we compute the limits set by LEP1
and predict the \dil\ coupling and mass ranges
which future high energy experiments 
can cover.

\section{Lagrangians}

In this Section, we construct the most general
lepton number conserving renormalisable lagrangian,
 consistent 
with electroweak symmetry, that
involves \sm\ gauge bosons, leptons and
the scalar or vector \dil s. The \dil-lepton-lepton
couplings, which we do not require to conserve
lepton flavour, are free parameters. They may
be very small, in which case \dil s would
have little effect on low-energy data. However,
the lowest dimensional couplings to the photon
and $Z^0$ are
always finite and sizable, so \dil s within
the kinematically allowed mass range can always be
produced at colliders. 

\subsection{Interactions with Leptons}

We consider the most general 
$SU(2)_L \otimes U(1)_Y$ invariant
dimension four 
lagrangian
coupling bosons to two leptons.  We require the
interactions to conserve lepton number, but
not lepton family number. We separate the lagrangian
into a part involving $L=0$ \dil s, and a part
involving $L=2$ \dil s:

\bea
\label{lag0}
{\cal L}_{L=0}
& = & 
g_1 \bar{\ell} \gamma_\mu \ell
{L_1^\mu}
+
\tilde g_1 \bar{e} \gamma_\mu e
{\tilde L_1^\mu}
\\\nonumber
& + &
\tilde g_2 \bar{\ell} e
{L_2}
+ \mbox{ h.c.}
\\\nonumber
& + & 
g_3 \bar{\ell} \gamma_\mu \vec\sigma \ell
\cdot \vec{L_3^\mu}
\\\nonumber\\
\label{lag2}
{\cal L}_{L=2}
& = & 
\lambda_1 \bar{\ell}^{{c}} i\sigma_2 \ell
{L_1} + \mbox{ h.c.}
+
\tilde \lambda_1 \bar{e}^{{c}} e
{\tilde L_1} + \mbox{ h.c.}
\\\nonumber
& + &
\lambda_2 \bar{\ell}^{{c}} \gamma_\mu e
{L_2^\mu} + \mbox{ h.c.}
\\\nonumber
& + &
\lambda_3 \bar{\ell}^{{c}} i\sigma_2\vec\sigma \ell
\cdot \vec{L_3} + \mbox{ h.c.}
~.
\nonumber
\eea

We have used the notation where 
$\ell = ( e_L, \nu_L)$ are left-handed $SU(2)_L$ lepton doublets 
and $e = e_R$ are right-handed charged singlet leptons.
The flavour indices are suppressed.
The charge conjugate fields in 
${\cal L}_{L=2}$
are defined to be
$\bar{\ell}^{{c}} = (\ell^c)^{\dagger} \gamma^o = -\ell^TC^{-1}$.
The subscript of the \dil\ fields $L_{1,2,3}$
indicates their $SU(2)_L$ 
singlet, doublet or triplet nature,
and the $\sigma$s are the Pauli matrices.
The vector \dil s also carry the Lorentz index $\mu$.
We list the quantum numbers of the 
\dil s in Table \ref{tqn}.

\newcommand{\sd}[1]{\raisebox{-2.5ex}[0ex][0ex]{$#1$}}
\renewcommand{\arraystretch}{2}

\begin{table}
$
\setlength{\arraycolsep}{1em}
\begin{array}{||c||c|c|c|c|c|c|l|c|}
  \hline
  \hline
  & L & J & Y & T_3 & Q_\gamma & Q_Z  &
  \multicolumn{1}{c|}{\mbox{lepton couplings}}
  & \mbox{familiar sibling} \\
  \hline
  \hline
L_1^\mu & 0 & 1 & 0 & 0 & 0 & 0 & \bar e_Le_L ~(g_1) 
  \quad \bar\nu_L\nu_L ~(g_1) & \gamma \quad Z^0 \quad Z' \\
  \hline
\tilde L_1^\mu & 0 & 1 & 0 & 0 & 0 & 0 & 
  \bar e_Re_R ~(\tilde g_1) & \gamma \quad Z^0 \quad Z' \\
  \hline
\sd{L_2} & \sd{0} & \sd{0} & \sd{1/2} & 1/2 & 1 &
 - {2\swt-1\over2\sw\cw} & \bar\nu_Le_R ~(g_2) & H^+ \\
  &&&& -1/2 & 0 &- {1\over2\sw\cw} & \bar e_Le_R ~(g_2) & H \\
  \hline
&&&& 1 & 1 & {\cw\over\sw} & \bar\nu_Le_L ~(\sqrt{2}g_3) 
  & W^+ \quad W'^+ \\
L_3^\mu & 0 & 1 & 0 & 0 & 0 & 0 & e_Le_L 
  ~(-g_3) \quad \bar\nu_L\nu_L ~(g_3) &\gamma \quad Z^0 \quad Z' \\
&&&& -1 &- 1 & -{\cw\over\sw} & \bar e_L\nu_L ~(\sqrt{2}g_3) 
  & W^- \quad W'^- \\
  \hline
L_1 & 2 & 0 & 1 & 0 & 1 & -{\sw\over\cw} & 
  e_L\nu_L ~(\lambda_1) ~(\mbox{antisymm.}) \\
  \cline{1-8}
\tilde L_1 & 2 & 0 & 2 & 0 & 2 & -2{\sw\over\cw} 
  & e_Re_R ~(\tilde\lambda_1) \quad(\mbox{symm.}) \\
  \cline{1-8}
  \sd{L_2^\mu} & \sd{2} & \sd{1} & \sd{3/2} & 1/2 & 2 &
   -{4\swt-1\over2\sw\cw} & e_Re_L ~(\lambda_2) \\
&&&& -1/2 & 1 & - {2\swt+1\over2\sw\cw} & e_R\nu_L ~(\lambda_2) \\
  \cline{1-8}
&&&& 1 & 2 & -{2\swt-1\over\sw\cw} & e_Le_L ~(\sqrt{2}\lambda_3) \\ 
L_3 & 2 & 0 & 1 & 0 & 1 & -{\sw\over\cw} 
  & e_L\nu_L~(\lambda_3) \qquad(\mbox{symm.}) \\
&&&& -1 & 0 & - {1\over\sw\cw} & \nu_L\nu_L ~(-\sqrt{2}\lambda_3) \\
  \cline{1-8}
\end{array}
$
\vspace{6.5cm}
\caption{
Major quantum numbers and couplings of the \dil s.
}
\label{tqn}
\end{table}

\renewcommand{\arraystretch}{1}

In principle derivative couplings could also be considered.
However,
these higher dimensional operators
would be  suppressed below the \dil mass scale.
We therefore  concentrate 
on the minimal lagrangians (\ref{lag0},\ref{lag2}).

The $L=0$ \dil s
are familiar fields,
as they resemble the electroweak gauge vectors 
and  the neutral and charged Higgs scalars.
Particles of this type
have been of great interest in
particle phenomenology
during the past decades,
and many  publications
have already been devoted to this topic.
In addition,
there is no compelling reason
to prevent these fields
from also coupling to quarks
\footnote{ Three of the four $L=2$ multiplets
contain a doubly-charged element, which can not
couple to quarks. Furthermore, if the $L=2$
\dil s also interact with quarks, they could
mediate fast proton decay.},
so we do not consider  the $L=0$ \dil s any further here.

We rewrite 
the  lagrangian for the $L=2$ \dil s (\ref{lag2}),
which are the ones we wish to study,  
with explicit electron $e$ and neutrino $\nu$ fields,
their flavour indices
$(i,j=1,2,3)$
and the helicity projectors
$P_{R/L} = 1/2(1\pm\gamma_5)$ .
The \dil\ superscript is
its electric charge:

\bea
\label{lag}
{\cal L}_{L=2}
&=&
\lambda_1^{ij} \quad L_1^{-} \quad 
\left( \bar e^c_i P_L \nu_j - \bar e^c_j P_L \nu_i \right) \\\nonumber
&+&
\tilde\lambda_1^{{ij}} \quad \tilde L_1^{--} 
\quad \bar e^c_i P_R e_j \\\nonumber
&+&
\lambda_2^{ij} \quad L_{2\mu}^{-} 
\quad \bar\nu^c_i \gamma^\mu P_R e_j \\\nonumber
&+&
\lambda_2^{ij} \quad L_{2\mu}^{--} 
\quad \bar e^c_i \gamma^\mu P_R e_j \\\nonumber
&-&
\sqrt{2}\lambda_3^{{ij}} \quad L_{3}^{0} 
\quad \bar\nu^c_i P_L \nu_j \\\nonumber
&+&
\lambda_3^{{ij}} \quad L_{3}^{-} 
\quad \left( \bar e^c_i P_L \nu_j + \bar e^c_j P_L \nu_i \right) \\\nonumber
&+&
\sqrt{2}\lambda_3^{{ij}} 
\quad L_{3}^{--} \quad \bar e^c_i P_L e_j \\\nonumber
&+&
\mbox{ h.c.}
~. 
\eea

If the scalar $L_3$ acquires a vacuum expectation value
it becomes the familiar $L=2$ triplet that appears
in left-right symmetric models and in 
the Gelmini-Roncadelli majoron model.
In this case, 
$L_3^{--}$ is then the well-studied doubly-charged Higgs 
\cite{HHG} and couples to a pair of 
like-sign $W$ bosons
via the lepton number violating vacuum expectation value.
These doubly-charged Higgs bosons
can be singly produced via $W$ fusion
in hadron collisions \cite{VD}.
Unfortunately, 
the production rates at LHC are so low
that only masses up to 1.2 TeV 
can be probed \cite{HMPR}.
In any case, the triplet vacuum expectation value is 
generically required to be small because of the smallness of $\rho-1$,
so  we will assume here that it is zero.

\subsection{Leptonic Couplings Matrices}

It is clear from the lagrangian (\ref{lag})
that the coupling matrix for $L_1$ 
 is  antisymmetric in flavour space,
whereas  $\tilde L_1$  and $L_3$ have flavour symmetric couplings.
The  isosinglet  $L_1$ corresponds to the antisymmetric
part of the cross product of two doublets, and the triplet
$L_3$  to the symmetric part.  
In contrast, 
the vector $L_2$ can have an arbitrary $3 \times 3$ coupling matrix.
For simplicity, 
we will assume that all
the coupling constant matrices are real.
This means we are neglecting $CP$ violation,
and our bounds are really constraints on
$| \lambda |$ rather than $ \lambda$. 
Constraints from $CP$ violation were included in the
analysis of \cite{DH,MRS,MSch}.

The aim of this paper is to discuss  \dil s
in a model independent way.  
We will  list the low-energy constraints that we derive
in Section 2 without making assumptions about the
structure of the coupling matrices $\lambda$ (except
in the case of $\mu \to e \gamma$, see section 3.1.4). However, 
most of the low-energy constraints
arise from  the non-observation 
of flavour violation  induced by off-diagonal matrix elements.
This makes the model-independent bounds
difficult to interpret (see Tables \ref{ler} and \ref{lertau} in
section 3.4).
We will therefore  also list the low-energy bounds
using three representative models of coupling matrices,
which we briefly review now. This provides clearer,
but assumption dependent, information.

\subsubsection{Flavour Diagonal Couplings}

One of the simplest choices
is a flavour diagonal coupling matrix 
with identical strength to all three flavours:
\beq
\label{diag}
\lambda^{ij}
=
\lambda\delta^{ij}
\equiv
\lambda
\left(
\barr{c@{\quad}c@{\quad}c}
1 & 0 & 0 \\ 0 & 1 & 0 \\ 0 & 0 & 1
\earr
\right)
~.
\eeq
This 
may be natural for a gauge \dil\
if there is no leptonic CKM matrix. 
It clearly does not apply to $L_1$, whose
coupling matrix is antisymmetric. 
In this model, 
there are fewer low-energy constraints, 
because they usually originate from
flavour changing processes.

\subsubsection{Flavour Democracy}

Another possible Ansatz,
which maximally mixes all three families,
is inspired from a universal Yukawa interactions model \cite{fd}:
\beq
\label{demo}
\lambda^{ij}
\equiv
\lambda
\left(
\barr{c@{\quad}c@{\quad}c}
1 & 1 & 1 \\ 1 & 1 & 1 \\ 1 & 1 & 1
\earr
\right)
~.
\eeq
For $L_1$, we will assume that
the ``flavour democratic''  coupling
matrix is of the form:
\beq
\label{demo'}
\lambda^{ij}
\equiv
\lambda
\left(
\barr{cc@{~}c}
~0 & ~1 & ~1 \\ -1 & ~0 & ~1 \\ -1 & -1 & ~0
\earr
\right)
~.
\eeq
This naturally leads to
lepton flavour violating processes
and the resulting bounds from low-energy experiments
are rather strict.

\subsubsection{Flavour Infiltration}

There is a mechanism that defines generations 
in approximately the same way as for $u$ and $d$ type quarks,
with small CKM mixing angles.
In Ref.~\cite{AHR},
it was suggested that this ``approximate flavour symmetry''
may be a property of the low energy effective theory derived
from any extension of the \sm . 
In this case, 
the dilepton-lepton coupling matrix would be of the form

\begin{comment}{
Another possible  ansatz for the form of
the \dil\ coupling constant matrix $\lambda$, is to
guess the relative size of
the off-diagonal
elements using ``approximate flavour symmetries'' \cite{AHR}.
In this ansatz,  
there is some unknown mechanism that defines
generations in approximately the same way
for all the fermions (an approximate symmetry). If
this mechanism worked perfectly, then all coupling constant
matrices  would be diagonal in generation space. 
This is not exactly the case in the quark
sector, but the CKM mixing angles are
small, so it is approximately  true. in Ref.~\cite{AHR} it is argued
that this property of the CKM matrix may be  a property
of the low-energy effective theory,
and that it can therefore be assumed to hold
for arbitrary extensions of the \sm.
\cite{AHR}  argue that
the mixing angles between
generations $i$ and $j$ ($i < j$) are of
order $m_i/m_j$, which
works reasonably well in the quark sector.
If we also assume this  for the lepton-\dil\ coupling matrix,
we have
}\end{comment}

\beq
\label{lap}
\lambda^{ij}
\equiv
\left(
\begin{array}{c@{~}c@{~}c}
\lambda^{1} & 
\frac{m_e}{m_{\mu}} (\lambda^{2} -\lambda^{1}) & 
\frac{m_e}{m_{\tau}} (\lambda^{3}-\lambda^{1}) 
\\
\frac{m_e}{m_{\mu}} (\lambda^{2}-\lambda^{1}) & 
\lambda^{2} & 
\frac{m_{\mu}}{m_{\tau}} (\lambda^{3}-\lambda^{2}) 
\\
\frac{m_e}{m_{\tau}} (\lambda^{3}-\lambda^{1}) & 
\frac{m_{\mu}}{m_{\tau}} (\lambda^{3}-\lambda^{2}) & 
\lambda^{3}
\end{array}
\right)
~.
\eeq

This structure of the coupling matrix,
which involves only three free parameters
may be particularly appropriate 
for the scalar \dil s. We anti-symmetrize this
matrix for $L_1$ (whose coupling matrix must
be anti-symmetric in generation space)
in the obvious way: we set the diagonal elements to zero,
and put negative signs in front of the lower triangle.

\subsection{Interactions with Neutral Gauge Bosons}

The interactions of the scalar and vector \dil s 
with the neutral gauge fields
are described by the following lagrangians:

\bea
\label{lagscal}
{\cal L}_{J=0} 
& = & 
{\left(D_\mu{L}\right)}^\dagger \left(D^\mu{L}\right)
\\\nonumber\\
\label{lagvec}
{\cal L}_{J=1} 
& = & 
-1/2 {\left(D_\mu{L^\nu}-D_\nu{L^\mu}\right)}^\dagger 
           \left(D^\mu{L_\nu}-D^\nu{L_\mu}\right)
\\\nonumber
&&-
i
(1-\kappa_\gamma)
eQ_\gamma {L^\dagger_\mu}{L_\nu} \left(\partial^\mu A^\nu - \partial^\nu A^\mu\right)
\\\nonumber
&&-
i
(1-\kappa_Z)
eQ_Z {L^\dagger_\mu}{L_\nu} \left(\partial^\mu Z^\nu - \partial^\nu Z^\mu\right)
~,
\eea

where $L$ and $L_\mu$ denote generic scalar and vector \dil s.
The covariant derivative is given by

\beq
\label{covder}
D_\mu = \partial_\mu - ie Q_\gamma A_\mu - ie Q_Z Z_\mu
~.
\eeq

The electric and weak charges 
$Q_\gamma$ and $Q_Z$
are uniquely determined by their hypercharge $Y$ 
and weak isospin projection $T_3$
\beq
\label{charges}
Q_\gamma = T_3 + Y
\qquad
Q_Z = T_3 {\cw \over \sw}
    - Y   {\sw \over \cw}
~.
\eeq
They are listed in Table~\ref{tqn}
along with the other quantum numbers
for all \dil s.

For the vector \dil s
there is an extra complication
due to our {\em a priori} ignorance 
of their gauge nature.
If they are gauge bosons
the {\em anomalous couplings} 
$\kappa_\gamma$ and $\kappa_Z$
in the lagrangian (\ref{lagvec})
vanish at tree level.
Finite values of these two parameters
would generate 
electric quadrupole and anomalous magnetic dipole moments
of the \dil s.
The {\em minimal couplings} are obtained for 
$\kappa_\gamma = \kappa_Z = 1$,
whereas for Yang-Mills \dil s
$\kappa_\gamma = \kappa_Z = 0$.
For simplicity,
we shall mostly focus on these two possibilities,
{\em i.e.},
$\kappa=0$ or $\kappa=1$.
Still, 
we keep in mind 
that in principle these parameters can take any value.

The \dil s may also have self-interactions 
or couplings to the Higgs or other  unobserved particles. 
Such interactions can in some cases be important, 
for instance in generating neutrino masses 
or magnetic moments \cite{CL,Z,LTV,FY}.
We do not consider this complication here,
as it involves the introduction of new free parameters
describing the couplings between \dil s and other bosons.
Without these interactions, the \dil s can only decay to a
pair of leptons.

\subsection{Mass Eigenstates}

Before proceeding further 
into any phenomenological analysis,
it is important to  note that the electroweak eigenstates
we have just described
may not be  mass eigenstates,
and that members of a given
multiplet may not have the same mass.
Those \dil s which carry the same spin 
and electromagnetic charge,
{\em i.e.},
$(L_1^-,L_3^-)$ and $(\tilde L_1^{--},L_3^{--})$,
could  mix. In principle,
if no particular model is referred to,
the mixing angles remain free parameters.
However
the couplings of the gauge eigenstates to leptons 
($\lambda_1,\tilde\lambda_1,\lambda_3$)
can easily be disentangled at colliders
by partitioning into symmetric and antisymmetric couplings
(for $L_1^-$ and $L_3^-$)
or with polarized experiments
(for $\tilde L_1^{--}$ and $L_3^{--}$).
We therefore shall not
 consider \dil\ mixing,
and we may use the couplings summarized in Table~\ref{tqn}
as such. 

The fact that members of a multiplet
may not have the same mass is relevant for
the low-energy bounds. 
If we assume that the \dil s acquire an $SU(2) \times U(1)$
invariant mass somewhere above
the electroweak scale, then any mass splittings between
members of an $SU(2)$ multiplet should be comparatively
small.  In this case, a bound on $\lambda^2/m^2$ for
one member of a multiplet applies
approximately to
other members. We nonetheless
list the bounds on multiplet members of different
electric charge separately, because  there is no
absolute guarantee  that the masses of $SU(2)$ multiplets are
approximately degenerate.

\section{Low-Energy Bounds}

Low-energy bounds on the \dil s 
can be derived  from the good agreement
between theory and experiment in processes
expected in the \sm, and  from the non-observation of reactions
which are  forbidden or suppressed in the  \sm.
Following
the Particle Data Book \cite{PDB}, which
lists upper bounds on most branching ratios at
90\% confidence level, we list our constraints  ``at 2 $\sigma$''. 

The mass of any $L=2$ \dil s is constrained to exceed at least 38 GeV
by LEP1 (see Section 4.1). 
This means that we can
usually approximate the low-energy effects
of \dil s in terms of four lepton operators. 
Details on how to derive the four fermion interactions
are presented in Appendix A.
The various renormalizable \dil\ interactions
and the four-fermion interactions that they induce
are listed in Table \ref{tff}.
We will use these repeatedly throughout
this Section.

As previously discussed, we neglect the phases of
the coupling constants. This means that for
flavour non-diagonal processes the signs in front of
the four-fermion vertices in Table \ref{tff} 
are irrelevant, and we neglect them. In the text,
we generically compute bounds from each process
on a four fermion vertex of specified tensor structure,
but with arbitrary coefficient $a \lambda^2/m_L^2$.
The factor ``$a$'' contains the possible factors of 2 and 1/2
(see Table \ref{tff}).
In the tables of results
we list the bounds for each \dil.

We compute bounds
from muon physics in the first subsection, and
from tau physics in the second. In the third subsection,
we briefly list other constraints.

{\arraycolsep1em
\renewcommand{\arraystretch}{2}
\begin{table}
$$
\begin{array}{||l|r|r||} \hline \hline
\tilde{L}_{1}^{--}  
& 
{\tilde{\lambda}_1^{\{ij\}} \tilde{\lambda}_1^{\{kl\}*} \over m_L^2} 
(\bar{e^c_i} P_R e_j) 
(\bar{e_l}P_L e^c_k)  
& 
{1\over2}
{\tilde{\lambda}_1^{\{ij\}} \tilde{\lambda}_1^{\{kl\}*} \over m_L^2} 
(\bar{e_k} \gamma^{\mu} P_R e_i) 
(\bar{e_l} \gamma_{\mu} P_R e_j) 
\\\hline
L_1^- 
&
4 
{\lambda_1^{[ij]} \lambda_1^{[kl]*} \over m_L^2}
( \bar{ e^c_i}  P_L \nu_j) 
( \bar{ \nu_l}  P_R e^c_k) 
&
2 
{\lambda_1^{[ij]} \lambda_1^{[kl]*} \over m_L^2}
( \bar{ e_k} \gamma^{\mu}  P_L e_i)
( \bar{ \nu_l} \gamma_{\mu} P_L \nu_j) 
\\\hline
L_{2\mu}^{--} 
& 
{\lambda_2^{ij} \lambda_2^{kl*} \over m_L^2}
( \bar{e^c_i} \gamma^\mu P_R e_j) 
( \bar{e_l} \gamma_\mu P_R e^c_k) 
&
-
{\lambda_2^{ij} \lambda_2^{kl*} \over m_L^2}
( \bar{e_k} \gamma^{\mu} P_L e_i) 
( \bar{e_l} \gamma_{\mu} P_R e_j)
\\\hline
L_{2\mu}^- 
& 
{\lambda_2^{ij} \lambda_2^{kl*} \over m_L^2}
( \bar{\nu^c_i} \gamma^\mu P_R e_j) 
( \bar{e_l} \gamma_\mu P_R \nu^c_k) 
&
-
{\lambda_2^{ij} \lambda_2^{kl*} \over m_L^2}
( \bar{\nu_k} \gamma^{\mu} P_L \nu_i) 
( \bar{e_l} \gamma_{\mu} P_R e_j)
\\\hline
L_3^{--}
& 
2 
{\lambda_3^{\{ij\}} \lambda_3^{\{kl\}*} \over m_L^2}
( \bar{ e^c_i} P_L e_j) 
( \bar{ e_l} P_R e^c_k) 
&
{\lambda_3^{\{ij\}} \lambda_3^{\{kl\}*} \over m_L^2}
( \bar{ e_k} \gamma^{\mu} P_L e_i)
( \bar{ e_l} \gamma_{\mu} P_L e_j) 
\\\hline
L_3^-
& 
4 
{\lambda_3^{\{ij\}} \lambda_3^{\{kl\}*} \over m_L^2}
( \bar{ e^c_i} P_L \nu_j) 
( \bar{ \nu_l} P_R e^c_k) 
&
2 
{\lambda_3^{\{ij\}} \lambda_3^{\{kl\}*} \over m_L^2}
( \bar{ e_k} \gamma^{\mu}  P_L e_i)
( \bar{ \nu_l} \gamma_{\mu} P_L \nu_j) 
\\\hline
L_3^0
& 
2 
{\lambda_3^{\{ij\}} \lambda_3^{\{kl\}*} \over m_L^2}
( \bar{ \nu^c_i} P_L \nu_j) 
( \bar{ \nu_l} P_R \nu^c_k) 
&
{\lambda_3^{\{ij\}} \lambda_3^{\{kl\}*} \over m_L^2}
( \bar{ \nu_k} \gamma^{\mu}  P_L \nu_i)
( \bar{ \nu_l} \gamma_{\mu} P_L \nu_j) 
\\\hline\hline
\end{array}
$$
\caption{
Four fermion vertices induced by the $L = 2$ \dil s, 
in their original and
Fierz-transformed
\sm-like forms.
The conversion is easily performed with the relations
(\ref{a1}--\ref{a3}) of Appendix A.
In the text
we assume the couplings are real. Coupling
constant indices in square brackets are
antisymmetric and  in curly brackets are symmetric.}
\label{tff}
\end{table}
\renewcommand{\arraystretch}{1}
}

\subsection{Muon Physics}

There are two categories of constraints from
muon physics. 
One can constrain the \dil s by requiring that their
contributions to decay modes forbidden in the \sm\
(by lepton family number conservation)
be less than the present experimental bounds.
One can also require that \dil\ contributions to
allowed \sm\  processes be ``sufficiently small''. 
This gives
a rough overall bound of  $\lambda^2/m_L^2 \lappeq G_F$,
that can in some cases be refined. 
For a clear (and entertaining) introduction to
muon physics, see Ref.~\cite{Scheck}.

\subsubsection{Polarized Muon Decay}

The influence of the singly-charged vector \dil\  $L^{\mu}_2$
on the decay of polarized muons 
has been discussed in Ref.~\cite{CF}.
If the four-fermion vertex mediating muon decay is a purely 
$(V-A) \times (V-A)$ effective interaction,
then when a polarized muon decays at rest, the electron spectrum
 has a particular shape. If the effective vertex
for the decay also contains, for instance, a  sufficiently
large  $(V - A)(V + A)$
contribution, then this can be detected in the
electron spectrum.

Following the notation of the Particle Data Book \cite{PDB}, 
we write  the most general four fermion vertex for muon
decay $\mu \to e \bar{\nu}_e \nu_{\mu}$,
 in terms of scalar, vector and tensor matrix
elements for left-handed and right-handed electrons and muons.
There are  experimental upper
bounds on the coefficient of each vertex,
(except of course the SM $(V-A) \times (V-A)$ vertex,
for which there is a lower bound). The coefficient $g^S_{RR}$ of
the vertex $ 2 \sqrt{2} G_F (\bar{e} P_L \nu)(\bar{\nu} P_R \mu)$
is required to be less than .066 \cite{J}.

By Fierz rearranging, we have
\beq
(\bar{e} \gamma^{\alpha}P_R \mu)(\bar{\nu}_{\mu} \gamma_{\alpha} P_L \nu_{e}) =
-2 (\bar{e} P_L \nu_{e})(\bar{\nu}_{\mu} P_R \mu) \label{ver}
~,
\eeq
so if $a \lambda^2/m_L^2$ is 
the coefficient  of the   $(V-A)(V+A)$ vertex (\ref{ver}), then
\beq
\frac{ a \lambda^2}{m_L^2} < g^S_{RR} \sqrt{2} G_F
\eeq
or
\beq
a \lambda^2 <  1.1 \left(\frac{m_L}{\rm TeV}\right)^2 \label{RHmu}
~.
\eeq
This applies to $L_{2\mu}^{-}$
and is listed in the tables as $\mu_R \to e \bar{\nu} \nu$.

\subsubsection{The Decay  $\mu^- \to e^- \bar{\nu}_{\mu} \nu_e$}

\begin{figure}[h]
\unitlength.5mm
\SetScale{1.418}
\begin{boldmath}
\begin{center}
\begin{picture}(60,40)(0,0)
\ArrowLine(0,20)(20,20)
\ArrowLine(40,30)(20,20)
\DashLine(20,20)(40,10){1}
\ArrowLine(40,10)(60,20)
\ArrowLine(40,10)(60,00)
\Text(-2,20)[r]{$\mu^-$}
\Text(42,30)[l]{$\bar\nu_\mu$}
\Text(62,20)[l]{$\nu_e$}
\Text(62,00)[l]{$e^-$}
\Text(30,10)[r]{$L^{-}$}
\end{picture}
\end{center}
\end{boldmath}
\caption{
Tree level process mediated by a singly-charged \dil\ 
that could induce $\mu^- \to e^-\nu_e\bar\nu_\mu$ decays.}
\label{fm2enn}
\end{figure}
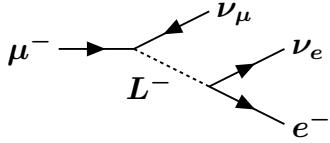

The  muon decay  $\mu^- \to e^- \bar{\nu}_{\mu} \nu_e$
has been considered in Refs~\cite{FN,FNS,FMSS,HM,CMS}.
This lepton flavour violating process is forbidden in the \sm, 
but can be mediated at tree level 
by the singly-charged \dil s
$L_{2\mu}^{-}$, $L_3^{-}$ and ${L}_1^{-}$
as depicted in Fig.~\ref{fm2enn}.
(Note that an $L=2$ \dil\ can only  conserve flavour 
if it couples to a single generation of leptons;
if it couples to multiple generations,
  its interactions can still be flavour diagonal, in the sense
that leptons of the same flavour meet at a vertex, as in
Eq.~(\ref{diag})). 
The experimental
bound on the branching ratio (compared
to ordinary muon decay) is B.R. $<$ 1.2\%  \cite{Krakauer91B}.
We assume that this applies to
 $(V-A)(V+A)$ vertices
as well as $(V-A)(V-A)$. 
There is also
a bound from the ratio \cite{S,11}
\beq
\frac{ \sigma(\bar{\nu}_{\mu} e^- \to \mu^- \bar{\nu}_e)}
{ \sigma({\nu}_{\mu} e^- \to \mu^- {\nu}_e)} \leq .05
\eeq
but we obtain a slightly better constraint from the decay:
\beq
\frac{a \lambda^2}{m_L^2} < \sqrt{.012} \times 2 \sqrt{2} G_F 
\eeq
or
\beq
a \lambda^2 < 3.6 \left( \frac{m_L}{\rm TeV} \right)^2
~.
\eeq
The results are listed in Table \ref{ler} in Section 3.4. This combination
of coupling constants will be better constrained by
the KARMEN neutrino oscillation experiment (see Section 3.3.3).

\subsubsection{The Decay $\mu \to 3e$}

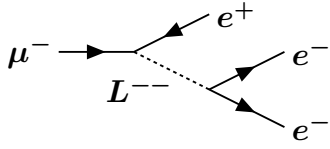
\begin{figure}[h]
\unitlength.5mm
\SetScale{1.418}
\begin{boldmath}
\begin{center}
\begin{picture}(60,40)(0,0)
\ArrowLine(0,20)(20,20)
\ArrowLine(40,30)(20,20)
\DashLine(20,20)(40,10){1}
\ArrowLine(40,10)(60,20)
\ArrowLine(40,10)(60,00)
\Text(-2,20)[r]{$\mu^-$}
\Text(42,30)[l]{$e^+$}
\Text(62,20)[l]{$e^-$}
\Text(62,00)[l]{$e^-$}
\Text(30,10)[r]{$L^{--}$}
\end{picture}
\end{center}
\end{boldmath}
\caption{
Tree level process mediated by a doubly-charged \dil\ 
that could induce $\mu^- \to e^-e^-e^+$ decays.}
\label{fm23e}
\end{figure}

The decay of a muon into three electrons 
has been discussed in Refs~\cite{DH,LTV,GGMKO}. 
This lepton flavour violating process is forbidden in the \sm, 
but can be mediated at tree level 
by the doubly-charged \dil s
$L_{2\mu}^{--}$, $L_3^{--}$ and $\tilde{L}_1^{--}$
as depicted in Fig.~\ref{fm23e}.
For $V\pm A$  \dil\ vertices, 
the rate ought to be the same
as for \sm\  muon decay (up to  electron
mass corrections), with $ a\lambda^2/m_L^2$ substituted for
$2 \sqrt{2} G_F$. The upper bound
on the branching ratio is $10^{-12}$ \cite{Bellgardt88}, 
so one has 
\beq
\sqrt{2} \frac{a \lambda^2}{m_L^2} < \sqrt{ 10^{-12}} \times 2 \sqrt{2} G_F 
\label{e17}
\eeq
or
\beq
a \lambda^2 < 2.3 \times 10^{-5} \left( \frac{m_L}{\rm TeV} \right)^2
~.
\eeq
The numerical bounds
are given in Table \ref{ler}, Section 3.4. 

Note that this {\em does} bound 4-fermion vertices
consisting of four fermions of the same chirality,
although at first sight this seems to imply making two
identical electrons at the same place. One can
see in the original vertex (before Fierz rearrangement) that
the two identical fermions do not multiply
to zero, but rather induce a Feynman rule vertex
$i 2 \lambda$. The rate in then divided by 2, for
identical fermions in the final state, which gives the
$\sqrt{2}$ on the left hand side of (\ref{e17}).

\subsubsection{The Decay $\mu \to e \gamma$}

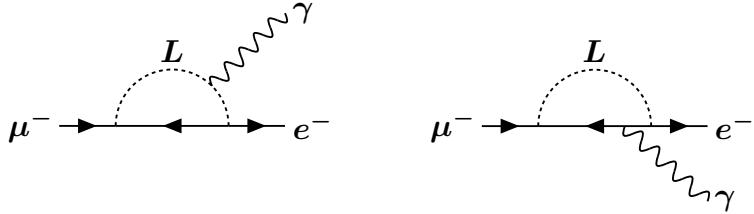
\begin{figure}[h]
\unitlength.5mm
\SetScale{1.418}
\begin{boldmath}
\begin{center}
\begin{picture}(60,40)(0,0)
\ArrowLine(0,0)(15,0)
\ArrowLine(45,0)(15,0)
\ArrowLine(45,0)(60,0)
\DashCArc(30,0)(15,0,180){1}
\Photon(40.6,10.6)(60,30){2}{5}
\Text(-2,0)[r]{$\mu^-$}
\Text(62,0)[l]{$e^-$}
\Text(62,30)[l]{$\gamma$}
\Text(30,20)[c]{$L$}
\end{picture}
\qquad\qquad\qquad
\begin{picture}(60,40)(0,20)
\ArrowLine(0,20)(15,20)
\ArrowLine(45,20)(15,20)
\ArrowLine(45,20)(60,20)
\DashCArc(30,20)(15,0,180){1}
\Photon(38,20)(60,0){-2}{5}
\Text(-2,20)[r]{$\mu^-$}
\Text(62,20)[l]{$e^-$}
\Text(62,0)[l]{$\gamma$}
\Text(30,40)[c]{$L$}
\end{picture}
\end{center}
\end{boldmath}
\vspace{10mm}
\caption{One-loop diagrams mediated by doubly- or singly-charged \dil s
that could induce $\mu \to e \gamma$.}
\label{fm2eg}
\end{figure}

The radiative muon decay has been
considered in Refs~\cite{Z,DH,LTV,CMS,BW}.
This lepton flavour violating process is forbidden in the \sm, 
but can be mediated at the one-loop level 
by the charged \dil s
as depicted in Fig.~\ref{fm2eg}.
The branching ratio of this decay is constrained to be very small:
$BR ( \mu \to e \gamma) <  4.9 \times 10^{-11}$ \cite{Bolton88}. 
However, this is  a one-loop process,
so the matrix element is suppressed by
a factor $ \sim (1/4 \pi)^2$. The decay
$\mu \to 3 e$ gives a stronger bound,
but $\mu \to e \gamma$ applies to 
different combinations of generation indices,
because one can have any lepton flavour in the loop. 

\begin{comment}{
The matrix element coupling two fermions and
an on-shell photon can be written 
\bea
&&{\cal M} [ f_i(p_i) \to f_f(p_f) + \gamma(q)] = 
\\\nonumber
&&- i \bar{u}_f(p_f) \left\{  
\frac{ \sigma^{\mu \nu} q_{\nu}}
{ m_i + m_f}  
\left[F_V(q^2) + F_A(q^2) \gamma_5\right] 
 +\gamma^{\mu}\left[ C_V(q^2)+ C_A(q^2) \gamma_5\right]\right\} u_i(p_i) \label{23}
~,
\eea
where $\sigma^{\mu \nu} \equiv \frac{-i}{2} [\gamma^{\mu}, \gamma^{\nu}]$,
and $(F_V + F_A)/(m_i + m_f)$  are the  magnetic ($F_V$) and electric
($F_A$) form factors. $C_V(0)$ and $C_A(0)$ are
zero by current conservation. 
}\end{comment}

The matrix element coupling two on-shell fermions to
an on-shell photon can be written 
\bea
{\cal M} [ f_i(p_i) \to f_f(p_f) + \gamma(q)] = 
\makebox[15em][l]{}
\\\nonumber
- i 
\bar{u}_f(p_f) 
\frac{ \sigma^{\mu \nu} q_{\nu}}
{ m_i + m_f}
\left[F_V(q^2) + F_A(q^2) \gamma_5\right] 
u_i(p_i) 
\label{23}
~,
\eea
where $\sigma^{\mu \nu} \equiv \frac{-i}{2} [\gamma^{\mu}, \gamma^{\nu}]$
and $q^2 = 0$. 
The decay rate of $f_i$ to $f_f + \gamma$ is 
\beq
\Gamma(f_i \to f_f + \gamma) = \frac{ m_i}{8 \pi} \left[F_V^2(0) +
F_A^2(0)\right]  \leq 1.5 \times 10^{-29} ~ {\rm GeV} \label{mudec}
~.\eeq
The  magnetic ($F_V(0)$) and electric
($F_A(0)$) dipole moments are both finite, and
$F_V \simeq F_A \equiv F$ for chiral fermions, so 
this gives 
\beq
F \leq 4.2 \times 10^{-14}
\eeq
for muon decay.

For scalar \dil s, 
we estimate the one-loop diagrams of Fig.~\ref{fm2eg} to be
\beq
F \simeq 
e \sum_k \left( \frac{Q_k}{12} - \frac{5 Q_L}{12} \right)
\frac{ \lambda^{\mu k} \lambda^{e k} m_{\mu}^2}{(4 \pi)^2 m_L^2} 
+ {\cal O} \left( \frac{ e\lambda^2 m_{\mu}^2 m_k^2}{( 4 \pi m_L^2)^2} 
 \right) 
 \label{sestL}
~,
\eeq
where $Q_L$ and $Q_k$ are the electric charges
of the \dil\ and the intermediate fermion.
For vector \dil s, 
we estimate the leading order contribution  
from the one-loop diagrams of Fig.~\ref{fm2eg}
to be
\beq
F \simeq 
e \sum_k \left( Q_k - \frac{3 Q_L}{4} \right)
\frac{  \lambda^{\mu k} \lambda^{e k} m_{\mu}^2}{(4 \pi)^2 m_L^2} 
 + {\cal O} \left( \frac{ e\lambda^2 m_{\mu}^2 m_k^2}{(4 \pi m_L^2)^2} 
\right) 
 \label{vestL}
~.
\eeq

We do not
need to flip the chirality
of the internal lepton, so we expect an even power of
the internal fermion mass.
Note that we have only estimated 
these matrix elements.
The constraints are
therefore only approximate, 
and could be missing factors of two.

Assuming there are no cancellations in the sum over internal leptons,
the leading order contribution in Eqs~(\ref{sestL},\ref{vestL}) dominates.
For the scalar \dil s
the approximate bound is then
\beq
a Q_{L}\sum_k \lambda^{\mu k} \lambda^{e k } \lappeq  4 \times 10^{-3} 
\left( \frac{m_L}{\rm TeV} \right)^2
\eeq
where $a=2$ if a pair of identical fermions meet
at a vertex, and 1 otherwise. 
The bound on the couplings of the vector
is of order
\beq
\left( Q_k - \frac{3 Q_L}{4} \right)
\sum_k \lambda^{\mu k} \lambda^{e k } \lappeq  2 \times 10^{-3} 
\left( \frac{m_L}{\rm TeV} \right)^2
~.
\eeq

Ignoring the sum over the internal fermions in Eqns (\ref{sestL})
and (\ref{vestL})
implies that
the bound on the sum 
is similar to the bound on the elements of the sum.
This is a non-trivial assumption; 
we do not expect this
to be the case for gauge vector bosons (for instance),
whose coupling matrices should be unitary. 
If, for any of the \dil s, the coupling matrices
are unitary ($ \sum_k \lambda^{\mu k} \lambda^{e k *} = 0$),
then the bounds we quote are overly optimistic, and
the first non-zero contribution to $\mu \to e \gamma$
is of order 
\beq
 \frac{\lambda^2}{(4 \pi)^2} \frac{ m_{\mu}^2 m_k^2}{m_L^4} ~.
\eeq
This does not give interesting bounds on the \dil s.
We therefore quote in the summary Table~\ref{ler} 
the ``non-GIM-suppressed'' 
bounds, while noting that they only apply
to \dil s whose coupling matrices are not unitary.
We explicitly consider the sum 
when we focus on specific models
in Tables~\ref{trdiag}--\ref{trafs}.

The product of couplings 
$\lambda^{\mu e} \lambda^{e e }$ is better
constrained by the $\mu \to 3e$ reaction.
In tables 3 and 4, we quote bounds on
$\lambda^{\mu \mu} \lambda^{\mu e }$ and
$\lambda^{\mu \tau} \lambda^{\tau e }$ 
assuming that the lowest order contributions
with different fermions in the loop do not
cancel against each other. 
As  discussed
in the previous paragraph, 
this may not be
a valid assumption.

As noted in Ref.~\cite{BW}, 
under certain circumstances
two-loop diagrams may give better bounds 
on the $SU(2)$ triplet \dil s than
the one-loop diagrams we have considered.
This  happens if the scalar develops a vacuum expectation value
and couples to the $W$ bosons;
as we explicitly ignore this possibility,
the one-loop diagram should dominate. 

\subsubsection{$g-2$ of the Muon}

The anomalous magnetic moment of the electron and the muon are two of
the most accurately measured quantities in
physics, and are frequently used to
constrain new particles. 
The measured value of
$(g-2)_{\mu} $ is \cite{PDB}
\beq
\frac{(g-2)_{\mu}}{2} = [11 659 230 \pm 84] \times 10^{-10} 
\eeq
and the \sm\ prediction \cite{Kin} is 
$(g-2)^{th}_{\mu} = 11 659 202 \pm 20$. This allows, at two $\sigma$,
a new physics contribution of order 
\beq
\delta \left( \frac{g-2}{2} \right)  = 170 \times 10^{-10} \label{g-2}
\eeq

The contribution to $g-2$ of
the muon from  \dil s,
 or generic bosons from beyond the \sm,
has been computed in  \cite{DH,BBKP,IKMMY,CK,FNS,FMSS,GGMKO,CMS,MWY}.

\begin{figure}[h]
\unitlength.5mm
\SetScale{1.418}
\begin{boldmath}
\begin{center}
\begin{picture}(60,40)(0,0)
\ArrowLine(0,0)(15,0)
\ArrowLine(45,0)(15,0)
\ArrowLine(45,0)(60,0)
\DashCArc(30,0)(15,0,180){1}
\Photon(40.6,10.6)(60,30){2}{5}
\Text(-2,0)[r]{$\mu^-$}
\Text(62,0)[l]{$\mu^-$}
\Text(62,30)[l]{$\gamma$}
\Text(30,20)[c]{$L$}
\end{picture}
\qquad\qquad\qquad
\begin{picture}(60,40)(0,20)
\ArrowLine(0,20)(15,20)
\ArrowLine(45,20)(15,20)
\ArrowLine(45,20)(60,20)
\DashCArc(30,20)(15,0,180){1}
\Photon(38,20)(60,0){-2}{5}
\Text(-2,20)[r]{$\mu^-$}
\Text(62,20)[l]{$\mu^-$}
\Text(62,0)[l]{$\gamma$}
\Text(30,40)[c]{$L$}
\end{picture}
\end{center}
\end{boldmath}
\vspace{10mm}
\caption{One-loop diagrams mediated by doubly- or singly-charged \dil s
that could contribute to $g-2$.}
\label{fg2}
\end{figure}
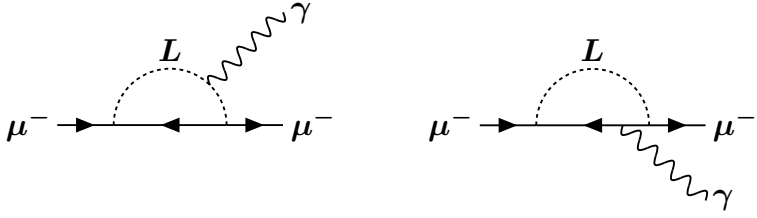

As we shall see, these constraints are not particularly
strong; we list them anyway, because unlike most low
energy bounds, $(g-2)/2$ constrains the square of a 
coupling constant. 

For the doubly-charged \dil s,
both diagrams in Fig.~\ref{fg2} contribute; for singly-charged
\dil s,
the photon does not couple to the
internal neutrino. These diagrams
have been evaluated in Refs~\cite{L,MWY}, from which we
can read off the leading order contributions.

If the \dil\ is a vector boson,
then the contribution to $(g-2)_{\mu}$ is
\beq
\left( \frac{2 Q_f}{3} -\frac{5 Q_L}{6}\right)
 \frac{h^2 m_{\mu}^2}{8 \pi^2 m_L^2}
\eeq
In these expressions and those that follow for scalar \dil s, 
$h$ is the coupling constant
appearing at the \dil-lepton-lepton vertex. It
can be read off from (\ref{lag}), providing one
remembers to include a factor of 2 at the vertices
where $L_1^{--}$ and $ L_3^{--}$ meet two identical
fermions.
For scalar \dil s, the contribution os
\beq
\left(\frac{Q_L}{12}-\frac{Q_f}{6} \right)  
\frac{h^2 m_{\mu}^2}{8 \pi^2 m_L^2} ~.
\eeq

From (\ref{g-2}), this gives bounds of order
\beq
\lambda^2 < 100 \left(\frac{m_L}{{\rm TeV}}\right)^2
\eeq
for the vector \dil s, and of order
\beq
\lambda^2 < 500 \left(\frac{m_L}{{\rm TeV}}\right)^2
\eeq
for the scalars. The exact bounds are listed in tables
\ref{ler} and \ref{lertau}. As noted earlier, these bounds are weak, but
apply to coupling constant combinations $ |\lambda^{\mu j} |^2$.
The bounds on \dil s from $g-2$ for the
electron are too weak to be interesting, because
the contributions  are suppressed by the electron mass squared.

\subsubsection{Muonium-Antimuonium Conversion}

\begin{figure}[h]
\unitlength.5mm
\SetScale{1.418}
\begin{boldmath}
\begin{center}
\begin{picture}(60,40)(0,0)
\ArrowLine(00,00)(30,00)
\ArrowLine(60,00)(30,00)
\ArrowLine(30,30)(00,30)
\ArrowLine(30,30)(60,30)
\DashLine(30,00)(30,30){1}
\Text(-2,0)[r]{$\mu^-$}
\Text(-2,30)[r]{$e^+$}
\Text(62,30)[l]{$e^-$}
\Text(62,0)[l]{$\mu^+$}
\Text(32,15)[l]{$L^{--}$}
\end{picture}
\end{center}
\end{boldmath}
\caption{
Muonium-antimuonium conversion 
mediated at tree-level by doubly-charged \dil s.}
\label{fMM}
\end{figure}
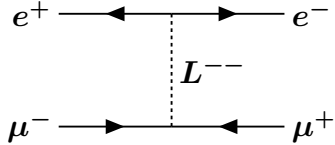

Doubly-charged \dil s with flavour diagonal couplings
may mediate 
the conversion of $\mu^+e^-$ atoms
(muonium)
into $\mu^-e^+$ atoms
(antimuonium)
as depicted in Fig.~\ref{fMM}. This has
been considered by many people
\cite{Pal,DH,BBKP,S,LTV,CK,FNS,FMSS,HS,H,HM}; 
\cite{FMSS} also studied muonium hyperfine 
splitting and \cite{HS} have studied this in 
a magnetic field.

As yet no such events have been observed
and the most stringent 90\%\ C.L. experimental bounds
on the effective 
$V \pm A$
interactions
are \cite{bertl}

\beq
\frac{a \lambda^2}{m_L^2}
<
\left\{
\barr{l@{\qquad}l}
0.018 \times 2 \sqrt{2} G_F & \mbox{for $(V+A)(V+A)$ or $(V-A)(V-A)$  vertices}
\\
0.012 \times 2 \sqrt{2} G_F & \mbox{for $(V+A)(V-A)$  vertices}
~.
\earr
\right.
\eeq
or
\beq
a \lambda^2 
<
\left\{
\barr{l@{\qquad}l}
.60 \left( \frac{m_L}{\rm TeV}\right)^2 & 
\mbox{for  $(V+A)(V+A)$ or $(V-A)(V-A)$  vertices}
\\
.40 \left(\frac{m_L}{\rm TeV} \right)^2 & 
\mbox{for   $(V+A)(V-A)$ vertices}
~.
\earr
\right.
\eeq
In the case of scalar \dil\ exchange, these are
identical fermions at the vertices, so the effective four-fermion
vertex is multiplied by 4.

\subsubsection{$G_F$ from Muon Decay}

Bilepton effects on the measurement of the Fermi constant
have been mentioned in Ref.~\cite{DH}, and discussed in Refs~\cite{Z,FMSS}.
$G_F$ is measured in muon decay to be 
$G_{\mu} = 1.16639(2) \times 10^{-5}$ \cite{PDB}. This is then
used to determine $V^{ud}$ in neutron $\beta$ decay.
As discussed in Ref.~\cite{MS}, one can constrain
new physics by comparing these leptonic and
hadronic determinations of $G$.
The \dil s could contribute at tree
level to muon decay, and 
make the experimental determination  of $
G_{\mu}$ larger or smaller than its ``true'' value $G_{\beta}$
that enters into  $\beta$ decay.
We assume that the \dil s
are the only new physics present,
so $G_{\beta}$ has the \sm\
value, and the CKM matrix is unitary.

It is clear that
\beq
G_{\mu} V_{ex} = G_{\beta} V_t \label{a}
\eeq
where $V_{ex}$ is the experimental determination
of a CKM angle, derived using $G_{\mu}$ for $G_F$, 
and $V_t$ is the ``true'' CKM matrix element.
Unitarity implies that
\beq
1 = \sum_i| V_t^{ui}|^2
\eeq
or, substituting from (\ref{a}) and rearranging
\beq
\frac{G_{\beta}^2}{G_{\mu}^2} = \sum_i |V_{ex}^{ui}|^2
\eeq
The experimental measurements
of $V_{ui}$ in the Particle
Data Book \cite{PDB}, at two sigma, imply
\beq
\sum_i| V_{ex}^{ui}|^2 = .9981 \pm .0055
\eeq
Setting $ 2 \sqrt{2} G_{\mu} = 2 \sqrt{2} G_{\beta} \pm a \lambda^2/m_L^2$,
we obtain 
\beq
-.0036< \frac{a S \lambda^2}{\sqrt{2} G_\beta m_L^2}
+ \frac{a^2 \lambda^4}{8 G_\beta^2 m_L^4} < .0074
\eeq
Note that we are allowing for interference
between the $W$ and the \dil; 
 $S$ is a possible chiral suppression factor that
arises if the \dil\ 
produces a right-handed rather than a left-handed electron. One
needs to flip the electron chirality  to interfere this
amplitude with the \sm\  one, so $S  =  m_e/ m_{\mu}$ \cite{FG}.
(See the Section on $G_F$ measured in $\tau$ decays for
a discussion of this, and more complete references.)
This gives
\beq
 -.059 \left( \frac{m_L}{\rm TeV} \right)^2 < a S \lambda^2 
< .12 \left( \frac{m_L}{\rm TeV} \right)^2 \label{39}
\eeq
for the interference term, and 
\beq
|a \lambda^2| < 
2.8 \left( \frac{m_L}{\rm TeV} \right)^2 
\label{G_FLL}
\eeq
for
the pure \dil\ contribution.  

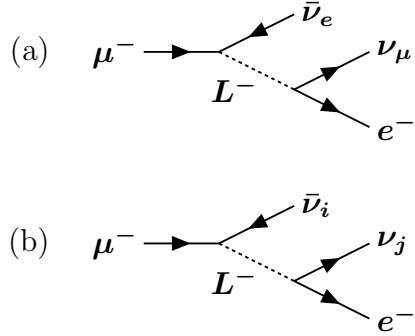
\begin{figure}[h]
\unitlength.5mm
\SetScale{1.418}
\begin{boldmath}
\begin{center}
\begin{picture}(60,40)(0,0)
\ArrowLine(0,20)(20,20)
\ArrowLine(40,30)(20,20)
\DashLine(20,20)(40,10){1}
\ArrowLine(40,10)(60,20)
\ArrowLine(40,10)(60,00)
\Text(-25,20)[r]{(a)}
\Text(-2,20)[r]{$\mu^-$}
\Text(42,30)[l]{$\bar\nu_e$}
\Text(62,20)[l]{$\nu_{\mu}$}
\Text(62,00)[l]{$e^-$}
\Text(30,10)[r]{$L^{-}$}
\end{picture}
\end{center}
~
~
\begin{center}
\begin{picture}(60,40)(0,0)
\ArrowLine(0,20)(20,20)
\ArrowLine(40,30)(20,20)
\DashLine(20,20)(40,10){1}
\ArrowLine(40,10)(60,20)
\ArrowLine(40,10)(60,00)
\Text(-25,20)[r]{(b)}
\Text(-2,20)[r]{$\mu^-$}
\Text(42,30)[l]{$\bar\nu_i$}
\Text(62,20)[l]{$\nu_j$}
\Text(62,00)[l]{$e^-$}
\Text(30,10)[r]{$L^{-}$}
\end{picture}
\end{center}
\end{boldmath}
\caption{
(a) Tree level process mediated by a singly-charged \dil\ 
that could interfere with the
standard model process  $\mu^- \to e^-\nu_e\bar\nu_\mu$.
(b) Pure bilepton contribution to $\mu \to e \bar{\nu} \nu$ 
(the neutrino flavours are arbitrary). }
\label{sacha}
\end{figure}

The interference
bound (\ref{39}) is very strong, because it
is a  bound on $G_F \lambda^2/m_L^2$, rather than
$\lambda^4/m_L^4$, and 
 unitarity is
satisfied at the $\sim .1$ \% level.
It applies to the scalar \dil s
$L_1^-$ and $L_3^{-}$.
with $(V-A)(V-A)$ vertices 
for the couplings mediating the decay of  
a $\mu_L^-$ to an  $e_L^-$, a $\nu_{\mu}$ and a $\bar{\nu}_e$.
This is a flavour non-diagonal process,
where the $\mu_L^-$ meets the $\bar{\nu}_e$ at the same vertex,
as shown in Fig.~\ref{sacha}a.

The bound on the purely \dil ic contribution (\ref{G_FLL})
constrains \dil\ decays to any flavour of neutrino,
as illustrated in Fig.~\ref{sacha}b.
The $(V+A)(V-A)$ vertices mediated by
the vector \dil\ $L_{2\mu}^-$, 
who suffer the helicity
suppression factor $S = m_e/m_{\mu}$, 
are better constrained
by the pseudoscalar matrix element constraint (\ref{RHmu}).

\subsubsection{The Weak Mixing Angle}

Bilepton effects on the measurement of
$\sin^2 \theta_W$  have  been considered in Ref.~\cite{CMS}.
The weak mixing angle is measured on the $Z^0$ resonance to
be $.2256 \pm .0023$ (on-shell scheme), 
and is measured leptonically at low E
in the ratio of $\nu_{\mu} e \to \nu_{\mu} e$ to
$\bar{\nu}_{\mu} e \to \bar{\nu}_{\mu} e$  to be .224 $\pm .009$.
Bileptons would contribute at tree level
to  the low-energy  determination of
$\sin^2 \theta_W$, and very little to the LEP
measurement, because LEP runs at the $Z^0$ peak.
 Bounds on \dil s
from $\nu_e e \to \nu_e e$  have also been calculated
\cite{FN};
these are similar to this, but constrain a different
combination of generation indices,  and will be discussed
in Section 3.3.4. 

Following the Particle Data Book notation, 
we write  the neutrino--electron
4-fermion vertex in the form
\beq
2 \sqrt{2} G_F (\bar{\nu} \gamma^{\mu} P_L \nu)
(\bar{e} \gamma_{\mu} [ g_L P_L + g_R P_R]e)
~,
\eeq
where 
$g_L = \frac{1}{2} (g_V - g_A),  g_R = \frac{1}{2} (g_V + g_A)$.
In the \sm\ one has at tree level
$g_L = \sin^2 \theta_W - 1/2, g_R = \sin^2 \theta_W$.
Using the
\sm\ tree level values for $g_L$ and $ g_R$, and 
$\delta g_{L,R} = a \lambda^2/(2 \sqrt{2} G_F m_L^2)$,
 the bound
\beq
\delta_{L,R} < 0.2256 -. (0.224 - 0.018)
\eeq
gives
\beq
|a \lambda^2| < 0.65  \left(\frac{m_L}{\rm TeV}\right)^2
\eeq
for $L_{2 \mu}^-$,
 $L_1^-$ and $L_3^-$.
These are bounds  on  the interference between
the \sm\ and \dil\ amplitudes, so they apply
to \dil\ vertices involving  $e$, $\bar{e}$, $\nu_{\mu}$ and
$\bar{\nu}_{\mu}$.

\subsection{Tau Physics}

The $\tau$ mass and
branching ratios have recently been measured
very accurately \cite{LP}, and the
new determinations are 
consistent within errors with the \sm.
The bounds
on new physics that can be derived from them
have been thoroughly and clearly discussed in Ref.~\cite{PS}. Here, we
make a more cursory analysis, but include
other rare decay bounds.

\subsubsection{$G_F$ from Tau Decay}

The  decays $\tau \to \ell \bar{\nu} \nu$ 
are  observed at 
the expected \sm\  rate,
which can be used to  constrain 
\dil s by requiring  that their
contributions to $G_F$ as measured in $\tau$ 
decays ($\equiv G_{\tau}$)  be sufficiently small. 
The neutrino flavour is unobserved,
so assuming that all the neutrinos are light,
we can use
this decay to bound
\dil -mediated decays
to arbitrary neutrino flavours. 

We assume that  there is no new physics
in the strongly interacting sector, so the
CKM matrix is unitary, and 
 $G_F$ in $\beta$ decay measures the four-fermion
vertex induced by the \sm\ $W$. 
From our previous discussion of the \dil\
contribution to $G$ in muon decay, we know
that at two sigma 
\beq
\frac{G_{\beta}}{\hat{G}_{\mu}}  = 1.0000 + .0018 - .0037
\eeq
where (see
\cite{PS,FG} for a complete analysis)
\beq
\hat{G}_{\mu}^2 \simeq G_{\mu}^2( 1 +  
2g_{RR}^S \frac{m_e}{m_{\mu}}) \label{Ghat}
\eeq
In our analysis of muon decay, we assumed that
$G_{\mu} = \hat{G}_{\mu}$, because $g_{RR}^S$ in
muon decay is
better constrained  by the shape of the electron
spectrum than by the size of $G$. (Recall that 
$\sqrt{2} g_{RR}^S G$ was the coefficient of
the $(V+A)(V-A)$ four fermion vertex (\ref{ver})). 
Eq.~(\ref{Ghat}) 
is a linear approximation; $g_{RR}^S$ appears
unsquared because it is multiplied by the
\sm\ vector coupling $ g_{LL}^V \simeq 1$. 

Recent $\tau$ data \cite{LP} implies, at one sigma,
\beq
\frac{\hat{G}_{e \tau}}{\hat{G}_{\mu}}  = .999 \pm .003
\eeq
and 
\beq
\frac{\hat{G}_{e \tau}}{\hat{G}_{\mu \tau}}  = 1.001 \pm .004
\eeq
where $G_{\ell \tau}$ is $G$ determined from
$\tau \to \ell \nu \bar{\nu}$ assuming a
$(V-A) \times (V-A)$ vertex.

Adding the errors
in these ``ratios of $G$'' in quadrature, we obtain 
at two sigma:
\beq
\frac{\hat{G}_{e \tau}}{G_{\beta}}  = 1.000 \pm .007 \label{Gex1}
\eeq
and
\beq
\frac{\hat{G}_{\mu \tau}}{G_{\beta}}  = 1.000 \pm .009  \label{Gex2}
\eeq
Alternatively, we could have constrained the ratio of
[$G$ that \dil s contribute to at
tree level]/[$G$ that \dil s do
not contribute to at tree level] from
the ratio $G_{\ell \tau}/G_{\pi, K}$. $G_{\ell \tau}$ is
$G$ determined from leptonic $\tau$ decays, and $G_{\pi, K}$
is $G$ determined from $\tau \to \pi \nu \bar{\nu}$
and $\tau \to K \nu \bar{\nu}$. The error in
this ratio is also of order 1-2\% \cite{PS,LP}, so it
gives the same constraints.

Bileptons can induce two types of
four fermion vertices that mediate tau decay: either
$(\bar{\ell} \gamma^{\mu} R \tau)(\bar{\nu} \gamma_{\mu} L \nu)$
or $(\bar{\ell} \gamma^{\mu} L \tau)(\bar{\nu} \gamma_{\mu} L \nu)$.
If   the right-handed
 vertex is present, then we have 
\beq
\hat{G}_{\ell \tau}^2 = G_{\beta}^2 \left( 1 + \frac{2  a \lambda^2}
{ \sqrt{2} m_L^2 G} \frac{m_{\ell}}{m_{\tau}} +  \frac{a^2 \lambda^4}
{8 m_L^4 G^2}
\right) 
\eeq
which implies, from the \sm-\dil\ interference term, that  
\beq
a\lambda^{3 2} \lambda^{23} < 1.9 \left(\frac{m_L}{\rm TeV}\right)^2 
\label{e51}
\eeq
 
If \dil s induce  the four-fermion vertex
$(\bar{\ell} \gamma^{\mu} L \tau)(\bar{\nu}_{\tau} \gamma_{\mu} L \nu_{\ell})$
then $\hat{G}$ would be
\beq
\hat{G}_{\ell \tau} = G_{\beta} \left( 1 + \frac{ a \lambda^2}
{2\sqrt{2} m_L^2 G} \right) 
\eeq
which gives the bound
\beq
a\lambda^{31} \lambda^{13} < .23  \left(\frac{m_L}{\rm TeV}\right)^2, ~~~
a\lambda^{32} \lambda^{23} < .32  \left(\frac{m_L}{\rm TeV}\right)^2
\eeq
on the singly-charged scalar \dil s 
$L_{1}^-$ and $ ~L_3^{-}$, if the
final states are such that there are interferences with the \sm\ amplitudes. 
For non-\sm\ processes similar to those depicted in Fig.~\ref{sacha}
the bounds are
\beq
a\lambda^{\nu 3} \lambda^{\nu' 1} < 2.7  \left(\frac{m_L}{\rm TeV}\right)^2, ~~~
a\lambda^{\nu 3} \lambda^{\nu' 2} < 3.3  \left(\frac{m_L}{\rm TeV}\right)^2
\eeq
for both $(V+A)(V-A)$ and $(V -A)(V-A)$ vertices.

\subsubsection{Lepton Flavour Violating Tau Decays}

There are also tau lepton bounds from the
non-observation of the flavour-changing
decays $\tau \to \ell\gamma$, and $\tau \to
3 \ell$ \cite{Hayes82,Keh88,Bowcock90}. 
Here (and in the tables) $\ell$ means $\mu$ or
$e$, and $f$ is any 
charged lepton or antilepton (so $3 \ell$ is
some $Q = 0$ combination of muons and
electrons). The bounds are straightforward copies
of the muon calculations, and the results
are listed in the tables. The experimental
upper bounds on the $\tau$ branching
ratios imply
\beq
F(\tau \to e \gamma) \lappeq 2.4 \times 10^{-8}  
\eeq
\beq
F(\tau \to \mu \gamma) \lappeq  4.0 \times 10^{-8}  
\eeq
which  give  bounds of order
\beq
Q \lambda^{\tau k} \lambda^{ e k} < 
 8 \left(\frac{m_L}{\rm TeV}\right)^2
\eeq
\beq
Q  \lambda^{\tau k} \lambda^{ \mu k}  < 
  13  \left(\frac{m_L}{\rm TeV}\right)^2 ~.
\eeq
We have approximated  $F_V \simeq F_A \equiv F$
as for the decay $\mu \rightarrow e \gamma$. 
We assume that the 
\dil\ coupling matrices $\lambda$ are
not unitary, so that their
contribution to $\tau \to \ell \gamma$ 
is not zero at lowest order (see the discussion
following Eq.~(\ref{vestL})).

The $\tau \to 3 \ell$ decays imply
\beq
a \lambda^2 <  .18 \left(\frac{m_L}{\rm TeV}\right)^2
\eeq
for the \dil s $~L_{2 \mu}^{--}, \tilde{L}_{1}^{--}$ and $ L_{3}^{--} $.
For simplicity, 
we conservatively neglect factors of 2 for identical fermions
in these bounds.

\subsection{Other Physics}

\subsubsection{Compositeness Searches}

Constraints on \dil s
from Bhabha scattering ($e^+ e^- \to e^+ e^-$)
have been calculated in
Refs~\cite{CK,FN,HSS,Gan}. However, 
 bounds on four fermion operators of the form
  $(\bar{e} \Gamma e)(\bar{f} \Gamma f)$ 
($f$ here  is an $e, \mu$ or $\tau$) have been
calculated by the JADE and TASSO collaborations
from PETRA data  \cite{Petra1,Petra2}, and 
it is these stronger constraints that we quote. 

It was pointed out
in Ref.~\cite{P}
 that composite theories would produce 
effective four fermion vertices of the
form
\bea
\frac{g^2 \eta_{LL}}{2 \Lambda^2} (\bar{\psi} \gamma^{\mu} P_L \psi) 
 (\bar{\psi} \gamma_{\mu} P_L \psi)  + 
\frac{g^2 \eta_{RR}}{2 \Lambda^2} (\bar{\psi} \gamma^{\mu} P_R \psi) 
 (\bar{\psi} \gamma_{\mu} P_R \psi)  \\ \nonumber
+
\frac{g^2 \eta_{RL}}{ \Lambda^2} (\bar{\psi} \gamma^{\mu} P_R \psi) 
 (\bar{\psi} \gamma_{\mu} P_L \psi) \label{ELP}
\eea
between the charged fermions. Including these
interactions in the cross-section
for $e^+ e^- \rightarrow e^+e^-, \mu^+ \mu^-, \tau^+ \tau^-$
changes the angular distribution of the final state particles. 
For $g^2 \simeq 4 \pi$, 
and $\eta_{PP'} = \pm 1$, JADE and TASSO give  bounds on $\Lambda$  
between  $1 - 7$ TeV \cite{Petra1,Petra2}.
 The exact numbers depend on the sign
and indices of $\eta$. Unfortunately,  $\eta_{LR}$,  does
not seem to have been explicitly considered, so we assume the bound
on $\Lambda$ in this case to be  $\simeq  3$ TeV. The exact constraints
are in the tables, but are roughly
\beq
a \lambda^2 \leq \frac{2 \pi m_L^2}{\Lambda^2}  \simeq .7 
\left(\frac{m_L}{\rm TeV} \right)^2
\eeq

Four fermion operators at LEP and SLC have been discussed
in Ref.~\cite{ECR}, who show that
experiments running on the $Z$ peak do not provide strong generic
bounds on four-fermion operators.

\subsubsection{Neutrino Oscillations}

Bileptons can be  constrained
by neutrino oscillation experiments if
the neutrinos are produced leptonically
(in the decay of a muon). In this case, the \dil s
could contribute at tree level to the production of
the neutrino beam, but are unimportant in the 
detection of the neutrinos, because the cross section
to scatter off an electron in the target is suppressed
with respect to the cross section on nuclei by a 
factor of order $m_e/E_{\nu}$. 
(Recall that \dil s do not couple to quarks.)
See Ref.~\cite{YW} for a thorough
discussion of constraints on various kinds of new
physics from neutrino oscillation experiments.

As usual, 
we assume that \dil s are the only source of physics
beyond the \sm.
They can mediate at tree level the
decays $\mu \to e \nu_i \bar{\nu}_j$ where
 $i \neq \mu$. The neutrino
oscillation experiment bounds on $ P_{\mu i} = 
\frac{1}{2} \sin^2 2 \theta$ for
large $\Delta m^2$  are upper limits on
the probability of making an $i$ neutrino when
one expected a $\mu$ neutrino.
  Most of these flavour oscillation
limits are of order $1-10 $\%, and therefore imply
weaker limits than can be derived from unitarity
arguments (section 3.1.6). However, as noted
in Ref.~\cite{YW}, the bounds from
the KARMEN experiment \cite{KARMEN} on $\nu_{\mu} \to \nu_e$
oscillations
and the FNAL E351  experiment \cite{FNAL} on $\nu_{\mu} \to \nu_{\tau}$
give 
\bea
P_{\mu e} < 3 \times 10^{-3} ~~ \\ \nonumber
P_{\mu \tau} < 2 \times 10^{-3} \label{nuosc}
~,
\eea
which translates into the bound
\beq
\frac{a \lambda^2}{m_L^2}  < \sqrt{P_{\mu i}} \times  2 \sqrt{2} G_F
\eeq
or
\bea
a\lambda^2 \lsim 1.8 \left( {m_L \over \mbox{TeV}} \right)^2 & 
 {\rm oscillation~to~electrons} \\
a\lambda^2 \lsim 1.5 \left( {m_L \over \mbox{TeV}} \right)^2 &
{\rm oscillation~to~taus}~.
\eea
This constraint applies to a \dil\ vertex involving
a $\mu$, $\bar{e}$, $\nu_f$ ($f$ is arbitrary) 
and a $\bar{\nu}_e$ or a $\bar{\nu}_{\tau}$.
Although the experimental results are very precise,
the ensuing bounds are not particularly strong
because there is no enhancement 
via interferences with a \sm\ amplitude.

\subsubsection{Neutrino Scattering}

The elastic scattering of neutrinos off electrons
is in principle a good place to look for generation diagonal bounds
on \dil s, but
 unfortunately the event rate is rather low. 
Constraints  from neutrino scattering processes
on vector \dil s
were considered in Ref.~\cite{FN}.

The $\nu_e e$ scattering cross-section in the
presence of right-handed neutrinos and an arbitrary four
fermion vertex was derived in Ref.~\cite{KFRS}. Assuming
that our neutrinos are left-handed, the vertex can be written
in terms of the effective coupling constants
$g_L = g_L^{SM} + \delta g_L $ and $g_R
= g_R^{SM} + \delta g_R $ (see Section 3.1.7).
These were measured in $\nu_e e$ scattering at LAMPF \cite{LAMPF},
to be \cite{Fol}
\bea
g_R^2 = .534 \pm .184 \\ \nonumber
g_L^2 = .084 \pm .031
~.
\eea
Using the LEP determination of $\sin^2 \theta_W$, we get
the two sigma bounds
\beq
{a \lambda^2} < 7~ \left( \frac{m_L^2}{{\rm TeV}} \right)^2
\eeq 
for 
 the singly-charged \dil s.

\subsubsection{Limits from Neutrino Masses and Magnetic Moments}

We have looked for  possible bounds on
\dil s from neutrino magnetic moments, but did
not find anything useful. 
We do not find bounds from neutrino
masses, because we assume that lepton
number is conserved, and that there are
no right-handed neutrinos.

\subsection{Low-Energy Summary}

In an attempt to clarify what ranges of
\dil\ mass and coupling constant are allowed,
we present the constraints in various ways.
We first simply list the bounds
in Table \ref{ler} and \ref{lertau}. 
This is not very appealing,
but does contain model-independent information. We
then make various assumptions about
the relative magnitudes of
the different entries in the coupling
constant matrices $\lambda^{ij}$,
as described in Section 2.2.
This allows
us to present attractive (but assumption
dependent) constraints in Tables
\ref{trdiag}, \ref{treq} and \ref{trafs}.

\renewcommand{\arraystretch}{2}
\begin{table}
$
\setlength{\arraycolsep}{.5em}
\begin{array}{||c||c|c|c|c|c|c||}
  \hline
  \hline
\mbox{experiment} & \tilde L_1^{--} & L_1^- & L_{2\mu}^{--} & L_{2\mu}^- & L_3^{--} & L_3^- 
\\\hline\hline
\mu_R \to e \bar\nu \nu
& 
& 
& 
& {\lambda^{1f} \lambda^{2f'} \over m_L^2} < 1  \times 10^{-0} 
& 
& 
\\\hline
\mu_L \to e \bar\nu_\mu \nu_e
& 
& 
& 
& {\lambda^{11} \lambda^{22} \over m_L^2} < 4 \times 10^{-0} 
& 
& {\lambda^{11} \lambda^{22} \over m_L^2} < 5 \times 10^{-1} 
\\\hline
G_\mu 
& 
& {\lambda^{[12]} \lambda^{[21]} \over m_L^2} < 6 \times 10^{-2} 
& 
&
& 
&  {\lambda^{12} \lambda^{21} \over m_L^2} < 6 \times 10^{-2} 
\\
& 
& {\lambda^{[1f]} \lambda^{[2f']} \over m_L^2} < 1 \times 10^{-0} 
& 
& {\lambda^{1f} \lambda^{2f'} \over m_L^2} < 3 \times 10^{-0} 
& 
&  {\lambda^{1f} \lambda^{2f'} \over m_L^2} < 1 \times 10^{-0} 
\\\hline
\sin^2\theta_w 
& 
& {\lambda^{[12]} \lambda^{[12]} \over m_L^2} < 3 \times 10^{-1} 
& 
& {\lambda^{12} \lambda^{12} \over m_L^2} < 6 \times 10^{-1} 
& 
& {\lambda^{12} \lambda^{12} \over m_L^2} < 3 \times 10^{-1} 
\\\hline
\mu \to e e \bar e 
& {\lambda^{11} \lambda^{12} \over m_L^2} < 5 \times 10^{-5} 
& 
& {\lambda^{11} \lambda^{12} \over m_L^2} < 2 \times 10^{-5} 
& 
& {\lambda^{11} \lambda^{12} \over m_L^2} < 2 \times 10^{-5} 
& 
\\\hline
\mu \to e \gamma 
& {\lambda^{f2} \lambda^{f1} \over m_L^2} < 2 \times 10^{-3} 
& {\lambda^{[f2]} \lambda^{[f1]} \over m_L^2} < 4 \times 10^{-3}
& {\lambda^{f2} \lambda^{f1} \over m_L^2} < 2 \times 10^{-3} 
& {\lambda^{f2} \lambda^{f1} \over m_L^2} < 2 \times 10^{-3}
& {\lambda^{f2} \lambda^{f1} \over m_L^2} < 1 \times 10^{-3}
& {\lambda^{f2} \lambda^{f1} \over m_L^2} < 2 \times 10^{-3}
\\
\hline
M \leftrightarrow \bar M 
& {\lambda^{11} \lambda^{22} \over m_L^2} < 3 \times 10^{-1} 
& 
& {\lambda^{11} \lambda^{22} \over m_L^2} < 4 \times 10^{-1} 
& 
& {\lambda^{11} \lambda^{22} \over m_L^2} < 2 \times 10^{-1} 
& 
\\\hline
(g-2)_{\mu} 
& {\lambda^{f2} \lambda^{f2} \over m_L^2} < 4 \times 10^{+2} 
& {\lambda^{[f2]} \lambda^{[f2]} \over m_L^2} < 1 \times 10^{+3} 
& {\lambda^{f2} \lambda^{f2} \over m_L^2} < 5  \times 10^{+1} 
& {\lambda^{f2} \lambda^{f2} \over m_L^2} < 1  \times 10^{+2} 
& {\lambda^{f2} \lambda^{f2} \over m_L^2} < 4  \times 10^{+2} 
& {\lambda^{f2} \lambda^{f2} \over m_L^2} < 1  \times 10^{+3} 
\\
\hline\hline
\end{array}
$
\medskip
\caption{}
\vspace{-.5ex}
\parbox{23cm}{
Summary of the
low-energy constraints on the \dil-lepton-lepton
coupling $\lambda$ from experiments
involving muons. 
The \dil\ mass is given in units of 1 TeV.
The bounds are only for absolute values of the couplings.
The bounds on the $L_2^\mu$ couplings
apply as well to the symmetrized combinations.
The labels $\ell$ stand for $e$ or $\mu$,
whereas the labels $f$ stand for all three families.
The $g-2$ bounds on $\lambda_1$ and $\lambda_3$
should be divided by 4 for $f = 2$; the $\mu \to e \gamma$ 
bounds on $\lambda_1$ and $\lambda_3$
should be divided by 2 for $f = 1,2$.  Both the
$g-2$ and $\mu \to e \gamma$  bounds apply to a sum
of coupling constant products; we assume  there are no
cancellations between terms in these bounds. }
\vspace{7.5cm}
\label{ler}
\end{table}

\renewcommand{\arraystretch}{2}
\begin{table}
$
\setlength{\arraycolsep}{.5em}
\begin{array}{||c||c|c|c|c|c|c||}
  \hline
  \hline
\mbox{experiment} & \tilde L_1^{--} & L_1^- & L_{2\mu}^{--} & L_{2\mu}^- & L_3^{--} & L_3^- 
\\\hline\hline
G_\tau 
& 
& {\lambda^{[3f]} \lambda^{[\ell f']} \over m_L^2} < 2 \times 10^{-0} 
& 
& {\lambda^{3f} \lambda^{ \ell f} \over m_L^2} < 3 \times 10^{-0} 
& 
& {\lambda^{3f} \lambda^{\ell f'} \over m_L^2} < 2 \times 10^{-0} 
\\
& 
& {\lambda^{[32]} \lambda^{[2 3]} \over m_L^2} < 2 \times 10^{-1} 
& 
& {\lambda^{32} \lambda^{2 3} \over m_L^2} < 2 \times 10^{-0} 
& 
& {\lambda^{32} \lambda^{2 3} \over m_L^2} < 2 \times 10^{-1} 
\\
& 
& {\lambda^{[3 1]} \lambda^{[1 3]} \over m_L^2} < 1 \times 10^{-1} 
& 
& 
& 
& {\lambda^{3 1} \lambda^{1 3} \over m_L^2} < 1 \times 10^{-1} 
\\\hline
\tau \to 3\ell 
& {\lambda^{3\ell} \lambda^{\ell'\ell''} \over m_L^2} < 4 \times 10^{-1} 
& 
& {\lambda^{3\ell} \lambda^{\ell'\ell''} \over m_L^2} < 2 \times 10^{-1} 
& 
& {\lambda^{3\ell} \lambda^{\ell'\ell''} \over m_L^2} < 2 \times 10^{-1} 
& 
\\\hline
\tau \to e \gamma 
& {\lambda^{f3} \lambda^{f1} \over m_L^2} < 4 \times 10^{-0} 
& {\lambda^{[f3]} \lambda^{[f1]} \over m_L^2} < 8 \times 10^{-0} 
& {\lambda^{f3} \lambda^{f1} \over m_L^2} < 4 \times 10^{-0} 
& {\lambda^{f3} \lambda^{f1} \over m_L^2} < 4 \times 10^{-0} 
& {\lambda^{3f} \lambda^{f1} \over m_L^2} < 4  \times 10^{-0} 
& {\lambda^{3f} \lambda^{f1} \over m_L^2} < 8  \times 10^{-0} 
\\\hline
\tau \to \mu \gamma 
& {\lambda^{f3} \lambda^{f2} \over m_L^2} <  7 \times 10^{-0} 
& {\lambda^{[f3]} \lambda^{[f2]} \over m_L^2} <  1 \times 10^{+1} 
& {\lambda^{f3} \lambda^{f2} \over m_L^2} <  7  \times 10^{-0} 
&  {\lambda^{f3} \lambda^{f2} \over m_L^2} <  7 \times 10^{-0} 
& {\lambda^{f3} \lambda^{f2} \over m_L^2} <  7 \times 10^{-0} 
& {\lambda^{f3} \lambda^{f2} \over m_L^2} <  1 \times 10^{+1}  
\\\hline
(\bar ee)(\bar ee)
& {\lambda^{11} \lambda^{11} \over m_L^2} < 2  \times 10^{-0} 
& 
& {\lambda^{11} \lambda^{11} \over m_L^2} < 7  \times 10^{-1}  
& 
& {\lambda^{11} \lambda^{11} \over m_L^2} < 8  \times 10^{-1}  
& 
\\\hline
(\bar ee)(\bar\mu\mu)
& {\lambda^{12} \lambda^{12} \over m_L^2} < 6  \times 10^{-1} 
& 
& {\lambda^{12} \lambda^{12} \over m_L^2} < 7  \times 10^{-1} 
& 
& {\lambda^{12} \lambda^{12} \over m_L^2} < 3  \times 10^{-1} 
& 
\\\hline
(\bar ee)(\bar\tau\tau)
& {\lambda^{13} \lambda^{13} \over m_L^2} < 3 \times 10^{-0} 
& 
& {\lambda^{13} \lambda^{13} \over m_L^2} < 7 \times 10^{-1} 
& 
& {\lambda^{13} \lambda^{13} \over m_L^2} < 1 \times 10^{-0} 
& 
\\\hline
\nu_{\mu} \to \nu_e, \nu_{\tau}
&
&
&
& {\lambda^{11} \lambda^{f2} \over m_L^2} < 1  \times 10^{-0} 
&
& {\lambda^{11} \lambda^{f2} \over m_L^2} < 3  \times 10^{-1}
\\
&
& {\lambda^{[13]} \lambda^{[f2]} \over m_L^2} < 9  \times 10^{-1}
&
& {\lambda^{13} \lambda^{f2} \over m_L^2} < 2 \times 10^{-0} 
&
& {\lambda^{13} \lambda^{f2} \over m_L^2} < 9  \times 10^{-1}
\\\hline
\nu_{e}e \to \nu_e e
&
&
&
& {\lambda^{11} \lambda^{11} \over m_L^2} < 7  \times 10^{-0} 
&
& {\lambda^{11} \lambda^{11} \over m_L^2} < 1  \times 10^{-0}
\\\hline \hline
\end{array}
$
\medskip
\caption{}
\vspace{-.5ex}
\parbox{23cm}{
Summary of the
low-energy constraints on the \dil-lepton-lepton
coupling $\lambda$ from $\tau$ decays, compositeness
searches and neutrino experiments.
The \dil\ mass is given in units of 1 TeV.
The bounds are only for absolute values of the couplings.
The bounds on the $L_2^\mu$ couplings
apply as well to the symmetrized combinations.
The labels $\ell$ stand for $e$ or $\mu$,
whereas the labels $f$ stand for all three families.
The  $\tau \to \ell \gamma$  bounds apply to a sum
of coupling constant products; we assume  there are no
cancellations between terms in these bounds. (The bounds on
$L^{--}_1$ and $L_3$ 
with arbitrary indices $f$ and/or $\ell$ should be
divided by 2 (or 4) if one (or both)  of the $\lambda$ is flavour diagonal.)}
\vspace{7.5cm}
\label{lertau}
\end{table}
\renewcommand{\arraystretch}{1}

In Tables \ref{ler} and \ref{lertau}, 
we list the
numerical bounds
on $\lambda^2 /m_L^2$ in units
of TeV$^{-2}$
for each  \dil\
from the various low-energy processes
we have considered.  
We list bounds separately for each member
of the \dil\ multiplets,
to allow 
different members of an
$SU(2)$ multiplet to have different masses. We assume,
in computing these bounds, that the \dil\ is the
only addition to the
\sm\, but
we make no assumptions about the
structure of the coupling constant
matrices.

The bounds listed in Tables \ref{ler} and \ref{lertau}  
are difficult to interpret. There are rarely constraints
on squares coupling
constant products of the form $|\lambda^{i j }|^2$. Rather,
 the bounds are usually on the products $\lambda^{i j} \lambda^{kl}$ with
$i \neq k, j \neq l$. There are two reasons for this. Firstly,
a number of constraints come from the non-observation of
lepton flavour violating interactions that
are forbidden in the \sm\, and usually these are
not mediated by generation diagonal \dil\ couplings. The
second difficulty is that most of
the data come from decays, which involve leptons of different
flavours for kinematic reasons. This means, for
example, that if there
was a $\sim 100$ GeV  boson with gauge strength
coupling only to muons or only to taus, 
we would not have seen it.

{\arraycolsep1em
\renewcommand{\arraystretch}{2}
\begin{table}
$$
\begin{array}{|| l | l | c ||} 
\hline\hline
\tilde{L}_1^{--} 
& {\lambda \over m_L} < .5~ \mbox{ TeV}^{-1} & M - \bar{M}
\\\hline
L_{2\mu}^{--} 
& {\lambda \over m_L} < .6 \mbox{ TeV}^{-1} & M - \bar{M}
\\\hline
L_{2\mu}^- 
& {\lambda \over m_L} < 1~ \mbox{ TeV}^{-1} & \mu_R \to e\nu\nu
\\\hline
L_3^{--} 
& {\lambda \over m_L} < .3  \mbox{ TeV}^{-1} & M - \bar{M}
\\\hline
L_3^- 
& {\lambda \over m_L} < .5 \mbox{ TeV}^{-1} & \nu_{\mu} \to \nu_e
\\\hline\hline
\end{array}
$$
\caption{
Best bounds on 
$\lambda / m_L$  
for each \dil,  
assuming flavour diagonal couplings (\protect\ref{diag}).
Note that these are bounds on $\lambda$, not $\lambda^2$.  
The processes from which the bounds originate
are listed in the third column.}
\label{trdiag}
\end{table}
\renewcommand{\arraystretch}{1}}

In Table \ref{trdiag}, 
we assume flavour diagonal interactions (\ref{diag}).
This  might
be a reasonable coupling constant matrix  for a gauge \dil, 
in the limit of massless neutrinos,
{\em i.e.},
in the absence of a leptonic CKM matrix.
The doubly-charged \dil s are best constrained by
the absence of muonium oscillations, 
whereas the singly-charged ones by neutrino oscillation
experiments and muon decay. 
New neutrino oscillation data could improve these bounds.

In Table \ref{treq}, 
we list the best constraint on each coupling constant
assuming flavour democracy (\ref{demo}).
The doubly-charged \dil s are best
constrained by the
non-observation of $\mu \to 3e$ decays,
whereas the singly-charged ones by radiative muon decays.

{\arraycolsep1em
\renewcommand{\arraystretch}{2}
\begin{table}
$$
\begin{array}{|| l | l | c ||} 
\hline\hline
\tilde{L}_1^{--} 
& {\lambda \over m_L} < 7 \times 10^{-3} \mbox{ TeV}^{-1} &  \mu \to 3e  
\\\hline
L_1^- 
& {\lambda \over m_L} < 6 \times 10^{-2} \mbox{ TeV}^{-1} &  \mu \to e \gamma  
\\\hline
L_{2\mu}^{--} 
& {\lambda \over m_L} < 4 \times 10^{-3} \mbox{ TeV}^{-1} &  \mu \to 3 e  
\\\hline
L_{2\mu}^- 
& {\lambda \over m_L} < 3 \times 10^{-2} \mbox{ TeV}^{-1} &  \mu \to e \gamma
\\\hline
L_3^{--} 
& {\lambda \over m_L} < 4 \times 10^{-3} \mbox{ TeV}^{-1} &  \mu \to 3e  
\\\hline
L_3^- 
& {\lambda \over m_L} < 2 \times 10^{-2} \mbox{ TeV}^{-1} &  \mu \to e \gamma  
\\\hline\hline
\end{array}
$$
\caption{
Best bounds on 
$\lambda / m_L$  
for each \dil,  
assuming flavour democracy (\protect\ref{demo}).
Note that these are bounds on $\lambda$, not $\lambda^2$.  
The processes from which the bounds originate
are listed in the third column.}
\label{treq}
\end{table}
\renewcommand{\arraystretch}{1}}

Assuming flavour infiltration (\ref{lap}),
we list in Table \ref{trafs}
the upper bounds on $\lambda^{1}, \lambda^{2}$ and $ \lambda^{3}$.
To compute bounds on the parameter $\lambda^{1}$
(for example), from a decay involving
$\lambda^{12} \lambda^{11}$, we set
$\lambda^{12} \simeq \frac{m_e}{m_{\mu}} \lambda^{1}$.
If $\lambda^{2} < \lambda^{1}$, this is a good
approximation, and if $\lambda^{2} > \lambda^{1} $,
it gives  conservative bounds. 
The strongest bounds originate from the non-observation 
of $\mu \to 3e$ decays 
and radiative muon and tau decays.

{\arraycolsep1em
\renewcommand{\arraystretch}{2}
\begin{table}
$$
\begin{array}{|| l | c|c|c ||} 
\hline\hline
\tilde{L}_1^{--} 
& {\lambda^1 \over m_L} < .1 \mbox{ TeV}^{-1} 
& {\lambda^2 \over m_L} < .6 \mbox{ TeV}^{-1} 
& {\lambda^3 \over m_L} < 11 \mbox{ TeV}^{-1} 
\\
& \mu \to 3e
& \mu \to e\gamma
& \tau \to \mu\gamma
\\\hline
L_1^- 
& {\lambda^1 \over m_L} < 50 \mbox{ TeV}^{-1} 
& {\lambda^2 \over m_L} < 8 \mbox{ TeV}^{-1} 
& {\lambda^3 \over m_L} < 8 \mbox{ TeV}^{-1} 
\\
& G_{\mu}
& G_{\tau}
& G_{\tau}
\\\hline
L_{2\mu}^{--} 
& {\lambda^1 \over m_L} < .06 \mbox{ TeV}^{-1} 
& {\lambda^2 \over m_L} < .6 \mbox{ TeV}^{-1} 
& {\lambda^3 \over m_L} < 11 \mbox{ TeV}^{-1} 
\\
& \mu \to 3e
& \mu \to e\gamma
& \tau \to \mu\gamma
\\\hline
L_{2\mu}^- 
& {\lambda^1 \over m_L} < .6 \mbox{ TeV}^{-1} 
& {\lambda^2 \over m_L} < .6~ \mbox{ TeV}^{-1} 
& {\lambda^3 \over m_L} <  11~ \mbox{ TeV}^{-1} 
\\
& \mu \to e \gamma
& \mu \to e\gamma
& \tau \to \mu\gamma
\\\hline
L_3^{--} 
& {\lambda^1 \over m_L} < .06 \mbox{ TeV}^{-1} 
& {\lambda^2 \over m_L} < .5 \mbox{ TeV}^{-1} 
& {\lambda^3 \over m_L} < 11 \mbox{ TeV}^{-1} 
\\
& \mu \to 3e
& \mu \to e \gamma
& \tau \to \mu \gamma
\\\hline
L_3^- 
& {\lambda^1 \over m_L} < .3 \mbox{ TeV}^{-1} 
& {\lambda^2 \over m_L} < .3 \mbox{ TeV}^{-1} 
& {\lambda^3 \over m_L} <  7 \mbox{ TeV}^{-1} 
\\
& \mu \to e \gamma
& \mu \to e\gamma
& \tau \to \mu\gamma
\\\hline\hline
\end{array}
$$
\caption{
Best bounds on 
$\lambda^i / m_L$  
for each \dil,  
assuming flavour infiltration (\protect\ref{lap}).
The superscript ``i'' is a generation index.
Note that these are bounds on $\lambda$, not $\lambda^2$.  
The processes from which the bounds originate
are listed below each bound.}
\label{trafs}
\end{table}
\renewcommand{\arraystretch}{1}}

To gauge the sensitivity of the low-energy experiments,
recall that the coefficient of
the $W$ mediated \sm\
4 fermion vertex is $2 \sqrt{2} G_F $, which,  in
our notation, would correspond to
$\lambda \simeq 6 m_L/$TeV.
The \sm\ Yukawa couplings for
leptons are of order $10^{-7} \to 10^{-2}$. One
can see from Tables \ref{trafs} and \ref{trdiag} 
that if one makes reasonable assumptions about the coupling
constant matrix $\lambda$, \dil s with masses
$\sim 100$ GeV $\to$ 10 TeV are consistent
with low-energy data.

\section{High Energy Bounds}

Since
\dil s do not couple to quarks,
the ideal high energy setup for their discovery 
seems to be a lepton and/or photon collider.
Nevertheless,
the Drell-Yan \dil\ pair-production mechanism
in hadron collisions
may become a possible source of \dil s
at LHC energies \cite{GMS}.
Of course,
even though by definition \dil s do not couple to quarks,
they may also carry a colour charge.
In this case hadron colliders may become a major source of \dil s,
but we ignore this exotic possibility here.

The present most stringent  bounds on \dil\ masses
originate from their non-observation in the $Z^0$ decays
\beq
\label{zdec}
Z^0 ~\to~ L^0L^0,~L^+L^-,~L^{++}L^{--}
\eeq
where $L$ represents a generic scalar or vector \dil.
As we shall see,
this constrains the \dil\ masses to lie above 38 GeV.
The advantage of these bounds
is that they are firm
and do not depend on any unknown lepton-\dil\ coupling.

Serious improvements on the present \dil\ bounds
are expected from experiments at a future linear collider,
with a typical energy in the TeV range.
A major asset of such a machine
is its versatility,
as it can be operated in the four \pe, \ee, \ep\ and \pp\ modes,
with highly polarized electron and photon beams.
The typical \lc\ designs
aim at an integrated yearly \pe\ luminosity
scaling with the squared \cm\ energy $s$
like
\beq
\label{lumpe}
{\cal L}_{e^+e^-} \mbox{ [fb$^{-1}$] } = 80 s \mbox{ [TeV$^2$] } 
\qquad \mbox{or} \qquad
{\cal L}_{e^+e^-} \simeq 3 \times 10^7 s
~.
\eeq
We shall present our results in such a way 
that departures from this working assumption
are trivial to implement.

For the doubly-charged \dil s
$\tilde L^{--}_1$, $L^{--}_{2\mu}$ and $L^{--}_3$,
indirect searches are possible in \pe\ and \ee\ scattering \cite{e-e-}
through the reactions:
\bea
\label{ipepe} \pe & ~\to~ & \pe
\\
\label{ipemm} \pe & ~\to~ & \ell^+\ell^-
\\
\label{ipepm} \pe & ~\to~ & e^+\ell^-
\\
\label{ieeee} \ee & ~\to~ & \ee
\\
\label{ieemm} \ee & ~\to~ & \ell^-\ell^-
\\
\label{ieeem} \ee & ~\to~ & e^-\ell^-
~,
\eea
where $\ell=\mu,\tau$.
The analysis of these processes
may significantly improve the excluded ratios
$\lambda/m_L$
obtained by low-energy data,
and provide additional information
on the structure of the coupling constants matrix.

In the event an anomaly is observed,
though,
it may as well be due to other ``new physics'' effects.
In this case,
such indirect evidence
is  no substitute for direct searches.

The lowest order reactions 
producing \dil s
are the following:
\bea
\label{deed} \ee & ~\to~ & L^{--}
\\
\label{dpes} \pe & ~\to~ & L^+~L^-
\\
\label{dped} \pe & ~\to~ & L^{++}~L^{--}
\\
\label{deps} \ep~ & ~\to~ & \bar\nu~L^-
\\
\label{depd} \ep~ & ~\to~ & e^+~L^{--}
\\
\label{dpps} \pp~~ & ~\to~ & L^+~L^-
\\
\label{dppd} \pp~~ & ~\to~ & L^{++}~L^{--}
~.
\eea

If the \cm\ energy reaches the mass 
of a doubly-charged \dil\
$\tilde L^{--}_1$, $L^{--}_{2\mu}$ and $L^{--}_3$,
a clearly outstanding resonance is expected from the reactions (\ref{deed}).
This \ee\ annihilation process 
obviously dwarfs the other reactions
(\ref{dped},\ref{depd},\ref{dppd}),
which therefore need not be considered.

Similarly,
the singly-charged \dil s
$L^-_1$ and $L^-_3$,
are obtained best in the \ep\ \lc\ operating mode
via the reactions (\ref{deps}).
Indeed,
this way the \dil s need not be pair produced as in reactions 
(\ref{dpes},\ref{dpps}),
and can be observed with lower \cm\ energies.

In the event
the lepton-\dil\ couplings are small,
though,
it may be difficult to resolve 
the \dil ic signal in the reactions (\ref{deps}).
In this case
the reactions (\ref{dpes},\ref{dpps})
offer an interesting alternative,
since they can proceed via 
the photon or $Z^0$ couplings to \dil s,
which always remain sizable.

As for low-energy experiments,
the most problematic \dil\
is the neutral $L^0_3$,
which does not couple to charged leptons
and will hence be more difficult to produce and to detect
in standard high energy experiments.
The only bounds 
which we can think of,
stem from the invisible $Z^0$ width.
If someday a $Z'$ resonance is reached,
a similar analysis will of course further constrain the $L^0_3$ mass.

\subsection{$Z^0$ Decay}

The tree-level $Z^0$ decay widths (\ref{zdec})
into a pair of scalar \cite{S,GMS,GGMKO} or vector \dil s
are given by:
\bea
\label{wscal}
\hspace{-2em}
\Gamma(J=0)
&=&
{\alpha Q_Z^2 \over 12} m_Z
\beta^3
\\
\label{wvec}
\hspace{-2em}
\Gamma(J=1)
&=&
{\alpha Q_Z^2 \over 12} m_Z
\beta^3
{1 \over 1-\beta^2}
\left[ 4\kappa^2+4\kappa-1 
     - 3\beta^2
     + 4(\kappa-1)^2 {1 \over 1-\beta^2} \right]
~,
\eea
where
\beq
\beta = \sqrt{1 - {4m_L^2 \over m_Z}}
\eeq
and $\kappa_Z$ is the weak anomalous coupling (\ref{lagvec}).

Since the four LEP experiments
report no serious deviation of the $Z^0$ width measurement
with its \sm\ prediction,
the \dil\  contributions (\ref{wscal},\ref{wvec})
to the charged lepton or invisible widths
may not exceed the experimental error,
which amount to 0.27 MeV and 4.2 MeV respectively
\cite{PDB}.

We do not know {\em a priori} the value of the anomalous weak coupling $\kappa_Z$
in Eq.~(\ref{wvec}).
The most conservative LEP bounds 
on vector \dil s
are given by the value of $\kappa_Z$
which minimizes the width
\beq
\label{kmin}
\kappa_{\rm min} = {1\over2} {1+\beta^2 \over 2-\beta^2}
~.
\eeq
The corresponding vector width is then given by
\beq
\label{wmin}
\Gamma_{\rm min}(J=1)
=
{\alpha Q_Z^2 \over 12} m_Z
\beta^3
{1 \over 1-\beta^2}
{ 5 - 4\beta^2 + 3\beta^4 \over 2 - \beta^2}
~.
\eeq

\begin{figure}[h]
\input{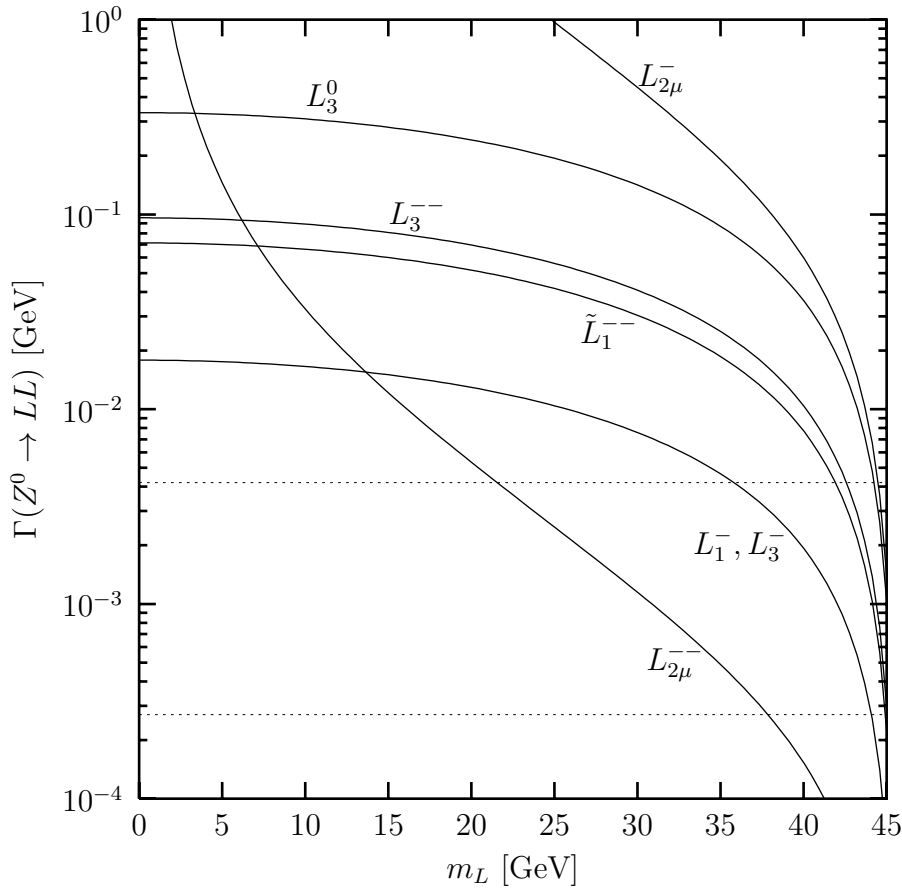}
\caption{
Partial decay widths of the $Z^0$ 
into pairs of \dil s
as a function of the \dil\ mass.
The lower dotted line shows 
the largest partial width consistent with data for the charged \dil s,
whereas the upper dotted line 
represents the same limit for the neutral \dil\ $L_3^0$.
}
\label{fzw}
\end{figure}

We plot $\Gamma_{\rm min}$ in Fig.~\ref{fzw}
as a function of the \dil\ mass.
The least constrained \dil\ turns out to be
$L_{2\mu}^{--}$,
whose coupling to the $Z^0$ vanishes in the limit
$\swt=1/4$.
Any mass above {\em ca} 38 GeV is allowed.
The other \dil s
$L_1^-,L_3^-,L_3^0$
must lie above 44 GeV
and
$\tilde L_1^{--},L_3^{--},L_{2\mu}^-$
above 45 GeV.

These results are confirmed 
by dedicated searches for doubly-charged Higgs bosons \cite{S',opal}.
Of course,
more specific dedicated searches
for four-lepton events
or highly ionizing tracks
(in the event the \dil\ couples so weakly to leptons
that it decays outside the detector)
may improve these bounds,
which anyway correspond to the worst case scenario
where $\kappa_Z$ is tuned as in Eq.~(\ref{kmin}).

Virtual \dil s may also enhance the four-lepton events rates,
if their couplings to leptons are large.
However,
the bounds obtainable this way 
cannot compete with those from low-energy data \cite{FNKY}.

\subsection{Indirect Signals}

The doubly-charged \dil s 
$\tilde L^{--}_1$, $L^{--}_{2\mu}$ and $L^{--}_3$
contribute to both Bhabha and M\o ller scattering \cite{e-e-},
even if they are too heavy to be directly produced
at the given collider energy.
They can therefore be detected 
via deviations from the \sm\ expectations
for the total \xs s and angular correlations.
In the presence of off-diagonal flavour couplings,
they may even produce final states
which are not expected in the realm of the \sm.

These processes have been considered previously
for the doubly-charged scalar \dil s 
in the context of triplet Higgs models \cite{R,S,GGMKO},
as well as for doubly-charged gauge \dil s \cite{FN,R'}.
Contrary to the claim of Ref.~\cite{R},
we do not find any chiral suppression in the scalar sector.

\subsubsection{\pe\ Scattering}

Doubly-charged \dil s contribute to Bhabha scattering (\ref{ipepe})
via their $u$-channel exchange
as depicted in Fig.~\ref{fpe}.
The corresponding polarized differential \xs s
in the case of the  exchange of the scalar $L_1^{--}$
are given by

\bea
\label{pes}
&&
{d\sigma(\pe\to\pe,\tilde L^{--}_1) \over dt} 
=
{4\pi\alpha^2 \over s^2}
\\\nonumber&&\hskip-2em
\left\{
[RR]
  \left[
    \left( \sum_i R_i^2 \left( {u \over s-m_i^2} + {u \over t-m_i^2} \right)
               - 2{\lambda^2 \over e^2} {u \over u-m_L^2}  \right)^2
  \right.
    + \left( \sum_i L_iR_i {t \over s-m_i^2} \right)^2
  \right]
\\\nonumber&&\hskip-2em
+
[LL]
  \left[
    \left( \sum_i L_i^2 \left( {u \over s-m_i^2} + {u \over t-m_i^2} \right)
               \right)^2
    + \left( \sum_i L_iR_i {t \over s-m_i^2} \right)^2
  \right]
\\\nonumber&&\hskip-2em
\left.
+
[LR]
    \left( \sum_i L_iR_i {s \over t-m_i^2} \right)^2
\right\}
~,
\eea
where 
$e$ is the charge of the electron,
$\alpha=e^2/4\pi$,
$\lambda=\lambda_{ee}$ is the generic diagonal coupling of the \dil\
and
$s$, $t$ and $u$ are the Mandelstam variables.
The dependence on the polarizations 
of the positron and electron beams
$P_\pm$
is contained in the factors
\bea
\label{pol}
[RR] & = & \displaystyle{1 + P_+ + P_- + P_+P_- \over 4}
\\\nonumber
[LL] & = & \displaystyle{1 - P_+ - P_- + P_+P_- \over 4}
\\\nonumber
[LR] & = & \displaystyle{1 - P_+P_- \over 2}
~.
\eea
The summation runs over $i=\gamma,Z^0$
and the \sm\ couplings of the photon and $Z^0$ boson
to left- and right-handed leptons
are given by
\beq
\label{sm}
\left\{
  \barr{l}
  \strut R_\gamma = 1\\
  \strut L_\gamma = 1\makebox(0,20)[b]{}
  \earr
\right.
\qquad\qquad
\left\{
  \barr{l}
  \strut R_{Z^0} = \displaystyle{ -\sin\theta_w \over \cos\theta_w} \\
  \strut L_{Z^0} = \displaystyle{1-2\sin^2\theta_w \over 2\sin\theta_w\cos\theta_w}~.\\
  \earr
\right.
\eeq

Similarly,
the contribution due to $L_3^{--}$
is given by the simple replacements
\beq
\label{pesp}
{d\sigma(L^{--}_3) \over dt} 
={d\sigma(\tilde L^{--}_1) \over dt} 
\left( \lambda \to \sqrt{2}\lambda , [LL] \leftrightarrow [RR] \right)
~.
\eeq

\begin{figure}[h]
\unitlength.5mm
\SetScale{1.418}
\begin{boldmath}
\hskip1em
\begin{picture}(80,40)(0,0)
\ArrowLine(0,0)(15,15)
\ArrowLine(15,15)(0,30)
\Photon(15,15)(45,15){2}{5}
\ArrowLine(60,30)(45,15)
\ArrowLine(45,15)(60,0)
\Text(-2,0)[r]{$e^-$}
\Text(-2,30)[r]{$e^+$}
\Text(62,0)[l]{$e^-$}
\Text(62,30)[l]{$e^+$}
\Text(30,23)[c]{$\gamma$,${Z^0}$}
\end{picture}
\qquad
\begin{picture}(80,40)(0,0)
\ArrowLine(0,0)(30,0)
\ArrowLine(30,0)(60,0)
\ArrowLine(30,30)(0,30)
\ArrowLine(60,30)(30,30)
\Photon(30,0)(30,30){2}{5}
\Text(-2,0)[r]{$e^-$}
\Text(-2,30)[r]{$e^+$}
\Text(62,0)[l]{$e^-$}
\Text(62,30)[l]{$e^+$}
\Text(34,15)[l]{$\gamma$,${Z^0}$}
\end{picture}
\qquad
\begin{picture}(0,40)(0,0)
\ArrowLine(0,0)(30,0)
\ArrowLine(60,0)(30,0)
\ArrowLine(30,30)(0,30)
\ArrowLine(30,30)(60,30)
\DashLine(30,0)(30,30){1}
\Text(-2,0)[r]{$e^-$}
\Text(-2,30)[r]{$e^+$}
\Text(62,0)[l]{$e^+$}
\Text(62,30)[l]{$e^-$}
\Text(34,15)[l]{$L^{--}$}
\end{picture}
%
\end{boldmath}
\caption{
  Lowest order Feynman diagrams contributing to $\pe\to\pe$ scattering.
  The exchanged doubly-charged \dil\ $L^{--}$ in the third diagram,
  can be either the scalars $\tilde L^{--}_1,L^{--}_3$ 
  or the vector $L^{--}_{2\mu}$.
}
\label{fpe}
\end{figure}

For the exchange of the vector \dil,
we find the differential \xs s

\bea
\label{pev}
&&
{d\sigma(\pe\to\pe,L^{--}_{2\mu}) \over dt} 
=
{4\pi\alpha^2 \over s^2}
\\\nonumber&&\hskip2em
\left\{
[RR]
  \left[
    \left( \sum_i R_i^2 \left( {u \over s-m_i^2} + {u \over t-m_i^2} \right) \right)^2
  \right.
\right.
\\\nonumber&&\hskip2em
    \hskip3em
    + \left( \left( \sum_i L_iR_i {t \over s-m_i^2} \right)^2
               - 2{\lambda^2 \over e^2} {t \over u-m_L^2} \sum_iL_iR_i {t \over s-m_i^2} \right)
\\\nonumber&&\hskip2em
  \left.
    \hskip3em
    + \left( \left( {\lambda^2 \over e^2} {s \over u-m_L^2} \right)^2
               - 2{\lambda^2 \over e^2} {s \over u-m_L^2} \sum_i L_iR_i {s \over t-m_i^2} \right)
  \right]
\\\nonumber&&\hskip2em
+
[LL]
  \left[
    \left( \sum_i L_i^2 \left( {u \over s-m_i^2} + {u \over t-m_i^2} \right) \right)^2
  \right.
\\\nonumber&&\hskip2em
    \hskip3em
    + \left( \left( \sum_i L_iR_i {t \over s-m_i^2} \right)^2
               - 2{\lambda^2 \over e^2} {t \over u-m_L^2} \sum_i L_iR_i {t \over s-m_i^2} \right)
\\\nonumber&&\hskip2em
  \left.
    \hskip3em
    + \left( \left( {\lambda^2 \over e^2} {s \over u-m_L^2} \right)^2
               - 2{\lambda^2 \over e^2} {s \over u-m_L^2} \sum_i L_iR_i {s \over t-m_i^2} \right)
  \right]
\\\nonumber&&\hskip2em
\left.
+
[LR]
  \left[
    \left( \sum_i L_iR_i {s \over t-m_i^2} \right)^2
  + \left( {\lambda^2 \over e^2} {t \over u-m_L^2} \right)^2
  \right]
\right\}
~.
\eea
The unpolarized
($P_+=P_-=0$)
differential \xs\ (\ref{pev})
coincides with the one given in Ref.~\cite{FN}
but disagrees with Ref.~\cite{R'}.

The \xs s for \pe\ annihilation into a pair of other leptons (\ref{ipemm})
is easily obtained from Eqs~(\ref{pes}) and (\ref{pev})
by suppressing the $\gamma,Z^0$ $t$-channel exchange
and replacing the diagonal coupling constants matrix element
$\lambda$ 
by the off-diagonal
$\lambda_{e\ell}$.

Similarly,
if there are simultaneously diagonal and off-diagonal \dil\ couplings,
the \xs\ for reaction (\ref{ipepm})
is obtained by keeping only the non-\sm\ $u$-channel contributions
and replacing $\lambda^2$ by
$\lambda_{ee}\lambda_{e\ell}$.

\subsubsection{\ee\ Scattering}

As depicted in Fig.~\ref{fee},
doubly-charged \dil s contribute to M\o ller scattering (\ref{ieeee})
via their $s$-channel exchange \cite{e-e-}.
The corresponding polarized differential \xs s
can of course be obtained from those in Bhabha scattering
by crossing symmetry.
In the case of the scalar exchange
they are given by

\bea
\label{ees}
&&
{d\sigma(\ee\to\ee,\tilde L^{--}_1) \over dt} 
=
{2\pi\alpha^2 \over s^2}
\\\nonumber&&\hskip2em
\left\{
[RR]
  \left[
    \sum_i R_i^2 \left( {s \over t-m_i^2} + {s \over u-m_i^2} \right)
         - 2{\lambda^2 \over e^2} {s \over s-m_L^2} 
  \right]^2
\right.
\\\nonumber&&\hskip2em
+
[LL]
    \left[ \sum_i L_i^2 \left( {s \over t-m_i^2} + {s \over u-m_i^2} \right)
         \right]^2
\\\nonumber&&\hskip2em
\left.
+
[LR]
  \left[
    \left( \sum_i L_iR_i {t \over u-m_i^2} \right)^2
  + \left( \sum_i L_iR_i {u \over t-m_i^2} \right)^2
  \right]
\right\}
~,
\eea
where 
we have used the same notations as in the previous section.

\begin{figure}[h]
\unitlength.5mm
\SetScale{1.418}
\begin{boldmath}
\begin{center}
\begin{picture}(80,40)(0,0)
\ArrowLine(0,0)(30,0)
\ArrowLine(30,0)(60,0)
\ArrowLine(0,30)(30,30)
\ArrowLine(30,30)(60,30)
\Photon(30,0)(30,30){2}{5}
\Text(-2,0)[r]{$e^-$}
\Text(-2,30)[r]{$e^-$}
\Text(62,0)[l]{$e^-$}
\Text(62,30)[l]{$e^-$}
\Text(34,15)[l]{$\gamma$,${Z^0}$}
\end{picture}
\quad
\raisebox{6.5mm}{+ crossed}
\qquad\qquad
\begin{picture}(70,40)(0,0)
\ArrowLine(0,0)(15,15)
\ArrowLine(0,30)(15,15)
\DashLine(15,15)(45,15){1}
\ArrowLine(45,15)(60,30)
\ArrowLine(45,15)(60,0)
\Text(-2,0)[r]{$e^-$}
\Text(-2,30)[r]{$e^-$}
\Text(62,0)[l]{$e^-$}
\Text(62,30)[l]{$e^-$}
\Text(30,23)[c]{$L^{--}$}
\end{picture}
\end{center}
\end{boldmath}
\caption{
  Lowest order Feynman diagrams contributing to $\ee\to\ee$ scattering.
  The exchanged doubly-charged \dil\ $L^{--}$ in the third diagram,
  can be either the scalars $\tilde L^{--}_1,L^{--}_3$ 
  or the vector $L^{--}_{2\mu}$.
}
\label{fee}
\end{figure}

Again,
as in the \pe\ case,
the \xs s for the $L^{--}_3$ exchange
are related to Eq.~(\ref{ees}) 
by the substitutions (\ref{pesp}).

Similarly,
for the exchange of the vector \dil,
we find the differential \xs s
\bea
\label{eev}
&&
{d\sigma(\ee\to\ee,L^{--}_{2\mu}) \over dt} 
=
{2\pi\alpha^2 \over s^2}
\\\nonumber&&\hskip2em
\left\{
[RR]
  s^2
  \left[
    \sum_i R_i^2 \left( {1 \over t-m_i^2} + {1 \over u-m_i^2} \right)
  \right]^2
\right.
\\\nonumber&&\hskip2em
+ 
[LL]
  s^2
  \left[
    \sum_i L_i^2 \left( {1 \over t-m_i^2} + {1 \over u-m_i^2} \right)
  \right]^2
\\\nonumber&&\hskip2em
+
[LR]
  \left[
    t^2 \left( \sum_i L_iR_i {1 \over u-m_i^2} 
             - {\lambda^2 \over e^2} {1 \over s-m_L^2} \right)^2
  \right.
\\\nonumber&&\hskip2em
    \hskip3em
\left.
  \left.
  + u^2 \left( \sum_i L_iR_i {1 \over t-m_i^2} 
             - {\lambda^2 \over e^2} {1 \over s-m_L^2} \right)^2
  \right]
\right\}
~.
\eea

The \xs s for the reactions (\ref{ieemm},\ref{ieeem}),
which produce leptons other than electrons 
are easily obtained from Eqs~(\ref{ees}) and (\ref{eev})
by keeping only the non-\sm\ $s$-channel contributions
and replacing $\lambda^2$ by either
$\lambda_{ee}\lambda_{\ell\ell}$ or
$\lambda_{ee}\lambda_{e\ell}$.

\subsubsection{Discovery Limits}

The Bhabha scattering data 
gathered below the $Z^0$ peak
has allowed to constrain the scalar \cite{S,GGMKO} and vector \cite{FN} \dil\
masses and coupling by
\beq
\label{petra}
{\tilde\lambda^{ee}_1 \over m_L} \lsim 4.4
\qquad
{\lambda^{ee}_2 \over m_L} \lsim 2.1
\qquad
{\lambda^{ee}_3 \over m_L} \lsim 3.1
\eeq
at the 95 \%\ confidence level.
A similar analysis of the LEP  data above the $Z$
could  improve these bounds, with sufficient luminosity.

To gauge the discovery potential of the reactions
(\ref{ipepe}--\ref{ieeem}),
it is instructive to work in the limits
\beq
\label{crlim}
|P|=1
\qquad\qquad
m^2_Z \ll s \ll m^2_L
\qquad\qquad
\sin^2\theta_w={1\over4}
~,
\eeq
where the propagators and the \sm\ $\gamma,Z^0$ couplings (\ref{sm})
simplify drastically.

The reactions (\ref{ipepm},\ref{ieemm},\ref{ieeem})
cannot take place in the realm of the \sm.
Given the spectacular nature of lepton flavour violation,
we estimate that five events should suffice to establish a discovery.
We therefore need an average number of 9.15
Poisson distributed events 
such that {\em at least} 5 events
are observed with 95\%\ probability. 

The average number of events is given by
\beq
\label{number}
N = {\cal L} \int dt {d\sigma \over dt}
~,
\eeq
where $\cal L$ is the integrated luminosity.
Because there will be some amount of anti-pinch at the interaction point
for the \ee\ mode,
we assume it
can only achieve half the luminosity of the \pe\ mode \cite{jim}:
\beq
\label{lumee}
{\cal L}_{e^-e^-} = {\cal L}_{e^+e^-}/2
~,
\eeq
where the \pe\ luminosity follows the energy scaling law (\ref{lumpe}).

For the reactions (\ref{ipepe},\ref{ipemm},\ref{ieeee}),
which do take place whether there are \dil s or not,
we compute the Cram\'er-Rao limit 
({\em cf.} Appendix B)
\beq
\label{cr}
\chi_\infty^2 = {\cal L} \int dt
{
  \left( \displaystyle{d\sigma(\lambda) \over dt} -
         \displaystyle{d\sigma(\lambda=0) \over dt} \right)^2
  \over
  \displaystyle{d\sigma(\lambda=0) \over dt}
}
~.
\eeq
Setting the value of the estimator $\chi^2_\infty$ equal to 3.84,
we obtain the 95\%\ confidence bounds on the parameter $\lambda$.

The particularly clean environment 
of the \pe\ and above all \ee\ collisions,
largely justifies here the neglect of systematic errors.

It is straightforward to obtain the following lower bounds 
on the observable values of the ratios
$\lambda/m_L$: 
\newcommand{\rb}[1]{\raisebox{3ex}[-3ex]
{$#1\makebox(0,0){\quad\rule[1.5ex]{.4mm}{35mm}}$}}
\begin{displaymath}
\barr{l@{\quad}c@{\quad}c@{\quad}ccrr}
\multicolumn{1}{c}{\mbox{reaction}} & \mbox{polarization} & \makebox[6em]{\dil} & \multicolumn{4}{c}{\mbox{bound}}
\\
& P_-=P_+=+1 & \makebox[6em]{$\tilde L^{--}_1$} & 
\displaystyle{\lambda_{ee}^2 \over m_L^2} & \ge &
.23
& \displaystyle\sqrt{16\pi\chi^2_\infty \over s{\cal L}_{e^+e^-}}
\\
\pe\to\pe \makebox(0,0){\quad\rule[1.5ex]{.4mm}{25mm}}
& P_-=P_+=-1 & \makebox[6em]{$L^{--}_3$} & 
\displaystyle{\lambda_{ee}^2 \over m_L^2} & \ge &
.11
& \displaystyle\sqrt{16\pi\chi^2_\infty \over s{\cal L}_{e^+e^-}}
\\
& P_-=P_+=\pm1 & \makebox[6em]{$L^{--}_{2\mu}$} & 
\displaystyle{\lambda_{ee}^2 \over m_L^2} & \ge &
.24
& \displaystyle\sqrt{16\pi\chi^2_\infty \over s{\cal L}_{e^+e^-}}
\earr
\end{displaymath}
\begin{displaymath}
\barr{l@{\quad}c@{\quad}c@{\quad}ccrr}
& P_-=P_+=+1 & \makebox[6em]{$\tilde L^{--}_1$} & 
\displaystyle{\lambda_{e\ell}^2 \over m_L^2} & \ge &
.22
& \displaystyle\sqrt{16\pi\chi^2_\infty \over s{\cal L}_{e^+e^-}}
\\
\pe\to\ell^+\ell^- \makebox(0,0){\quad\rule[1.5ex]{.4mm}{25mm}}
& P_-=P_+=-1 & L^{--}_3 & 
\displaystyle{\lambda_{e\ell}^2 \over m_L^2} & \ge &
.11
& \displaystyle\sqrt{16\pi\chi^2_\infty \over s{\cal L}_{e^+e^-}}
\\
& P_-=P_+=\pm1 & L^{--}_{2\mu} & 
\displaystyle{\lambda_{e\ell}^2 \over m_L^2} & \ge & 
.53
& \displaystyle\sqrt{16\pi\chi^2_\infty \over s{\cal L}_{e^+e^-}}
\earr
\end{displaymath}
\begin{displaymath}
\barr{l@{\quad}c@{\quad}c@{\quad}ccrr}
& P_-=P_+=+1 & \makebox[5em]{$\tilde L^{--}_1$} & 
\displaystyle{\lambda_{ee} \over m_L}{\lambda_{e\ell} \over m_L} & \ge &
.31
& \displaystyle\sqrt{16\pi N \over s{\cal L}_{e^+e^-}}
\\
& P_-=P_+=-1 & L^{--}_3 & 
\displaystyle{\lambda_{ee} \over m_L}{\lambda_{e\ell} \over m_L} & \ge &
.15
& \displaystyle\sqrt{16\pi N \over s{\cal L}_{e^+e^-}}
\\
\rb{\pe\to e^+\ell^-}
& P_-=P_+=\pm1 & L^{--}_{2\mu} & 
\displaystyle{\lambda_{ee} \over m_L}{\lambda_{e\ell} \over m_L} & \ge &
.50
& \displaystyle\sqrt{16\pi N \over s{\cal L}_{e^+e^-}}
\\
& P_-=-P_+=\pm1 & L^{--}_{2\mu} & 
\displaystyle{\lambda_{ee} \over m_L}{\lambda_{e\ell} \over m_L} & \ge &
.87
& \displaystyle\sqrt{16\pi N \over s{\cal L}_{e^+e^-}}
\earr
\end{displaymath}
\begin{displaymath}
\barr{l@{\quad}c@{\quad}c@{\quad}ccrr}
& P_1=P_2=+1 & \makebox[6em]{$\tilde L^{--}_1$} & 
\displaystyle{\lambda_{ee}^2 \over m_L^2} & \ge &
.18
& \displaystyle\sqrt{16\pi\chi^2_\infty \over s{\cal L}_{e^+e^-}}
\\
\ee\to\ee \makebox(0,0){\quad\rule[1.5ex]{.4mm}{25mm}}
& P_1=P_2=-1 & L^{--}_3 & 
\displaystyle{\lambda_{ee}^2 \over m_L^2} & \ge &
.09
& \displaystyle\sqrt{16\pi\chi^2_\infty \over s{\cal L}_{e^+e^-}}
\\
& P_1=-P_2=\pm1 & L^{--}_{2\mu} & 
\displaystyle{\lambda_{ee}^2 \over m_L^2} & \ge &
.44
& \displaystyle\sqrt{16\pi\chi^2_\infty \over s{\cal L}_{e^+e^-}}
\earr
\end{displaymath}
\begin{displaymath}
\barr{l@{\quad}c@{\quad}c@{\quad}ccrr}
& P_1=P_2=+1 & \makebox[5.5em]{$\tilde L^{--}_1$} & 
\displaystyle{\lambda_{ee} \over m_L}{\lambda_{\ell\ell} \over m_L} & \ge &
.35
& \displaystyle\sqrt{16\pi N \over s{\cal L}_{e^+e^-}}
\\
\ee\to\ell^-\ell^- \makebox(0,0){\quad\rule[1.5ex]{.4mm}{25mm}}
& P_1=P_2=-1 & L^{--}_3 & 
\displaystyle{\lambda_{ee} \over m_L}{\lambda_{\ell\ell} \over m_L} & \ge &
.18
& \displaystyle\sqrt{16\pi N \over s{\cal L}_{e^+e^-}}
\\
& P_1=-P_2=\pm1 & L^{--}_{2\mu} & 
\displaystyle{\lambda_{ee} \over m_L}{\lambda_{\ell\ell} \over m_L} & \ge &
.87
& \displaystyle\sqrt{16\pi N \over s{\cal L}_{e^+e^-}}
\earr
\end{displaymath}
\begin{displaymath}
\barr{l@{\quad}c@{\quad}c@{\quad}ccrr}
& P_1=P_2=+1 & \makebox[5.5em]{$\tilde L^{--}_1$} & 
\displaystyle{\lambda_{ee} \over m_L}{\lambda_{e\ell} \over m_L} & \ge &
.25
& \displaystyle\sqrt{16\pi N \over s{\cal L}_{e^+e^-}}
\\
\ee\to e^-\ell^- \makebox(0,0){\quad\rule[1.5ex]{.4mm}{25mm}}
& P_1=P_2=-1 & L^{--}_3 & 
\displaystyle{\lambda_{ee} \over m_L}{\lambda_{e\ell} \over m_L} & \ge &
.13
& \displaystyle\sqrt{16\pi N \over s{\cal L}_{e^+e^-}}
\\
& P_1=-P_2=\pm1 & L^{--}_{2\mu} & 
\displaystyle{\lambda_{ee} \over m_L}{\lambda_{e\ell} \over m_L} & \ge &
.61
& \displaystyle\sqrt{16\pi N \over s{\cal L}_{e^+e^-}}
\earr
\end{displaymath}

If the positrons cannot be polarized,
the \pe\ bounds are worsened by a factor $\sqrt{2}$.

The bounds at 95\%\ confidence level
are obtained by setting 
$\chi^2_\infty=3.84$
and
$N=9.15$.
Note that the reactions which are allowed or forbidden
within the realm of the \sm\
all yield similar results.
Indeed,
although the allowed reactions need much more anomalous events
to be statistically relevant,
this number of anomalous events is very much enhanced 
by the interferences with the \sm\ channels.

\subsection{Direct Signals}

If the \cm\ energy of the collider is sufficient to produce \dil s,
their decay widths becomes an issue of importance.
As we ignore self-interactions
and assume the scalars do not develop a vacuum expectation value,
the \dil s cannot decay weakly
into a single or a pair of gauge bosons.
Therefore,
the leptonic two-body decay modes are dominant.
Setting all lepton masses to zero,
the scalar \dil\ widths are
\beq
\label{w}
\Gamma = A {m_L \over 8\pi} \sum_{ij} \lambda_{ij}^2
~,
\eeq
where $A=1/2,1,1/3,2$ 
for $L_1,~\tilde L_1,~L_{2\mu},~L_3$ respectively
and the sum runs over all elements of the coupling constants matrix
($i,j=e,\mu,\tau$).
The doubly-charged vector width 
agrees with the width computed in Ref.~\cite{FN}
and disagrees with Ref.~\cite{R'},
while the doubly-charged scalar widths
agree with Refs~\cite{S,GMS,GGMKO}
and disagree with Ref.~\cite{R}.

If the indirect searches 
from experiments below the real production threshold
have been unsuccessful,
the leptonic couplings are unlikely to exceed 0.2
and one can safely state that the total \dil\ width $\Gamma_L$ is narrow:
\beq
\label{width}
\Gamma_L \le 10^{-2} m_L
~.
\eeq
Of course,
this argumentation does not directly apply 
to the neutral and singly-charged \dil s
and may break down
for some perverse choices of coupling matrices.
We ignore this possibility here.

\subsubsection{\ee\ Scattering}

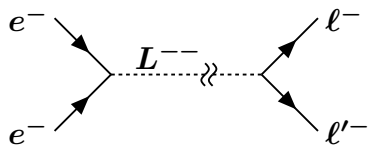
\begin{figure}[h]
\unitlength.5mm
\SetScale{1.418}
\begin{boldmath}
\begin{center}
\begin{picture}(70,40)(0,0)
\ArrowLine(0,30)(15,15)
\ArrowLine(0,0)(15,15)
\DashLine(15,15)(40,15){1}
\Photon(40,12)(40,18){1}{1}
\Photon(42,12)(42,18){1}{1}
\DashLine(42,15)(55,15){1}
\ArrowLine(55,15)(70,0)
\ArrowLine(55,15)(70,30)
\Text(-2,0)[r]{$e^-$}
\Text(-2,30)[r]{$e^-$}
\Text(30,20)[c]{$L^{--}$}
\Text(72,30)[l]{$\ell^-$}
\Text(72,0)[l]{$\ell'^-$}
\end{picture}
\end{center}
\end{boldmath}
\caption{
  Lowest order Feynman diagram
  for the production
  of a doubly-charged \dil\ $L^{--}$
  in \ee\ collisions.
  The produced \dil\
  can be either a scalar $\tilde L^{--}_1,L^{--}_3$ 
  or the vector $L^{--}_{2\mu}$.
}
\label{feeres}
\end{figure}

As depicted in Fig.~\ref{feeres},
a doubly-charged \dil\ can be produced in \ee\ collisions (\ref{deed})
and subsequently decay into a pair of leptons \cite{e-e-}.
The corresponding polarized \xs s
for the production and decay of scalars or vectors
is obtained by replacing the $s$-channel propagator
in Eqs~(\ref{ees},\ref{eev}) 
by a Breit-Wigner resonance
with the correct width (\ref{w}).
On the resonance
$s=m_L^2$,
we may ignore other possible channels
and find the \xs s
\bea
\label{xss}
\sigma(\ee\to\ell^-\ell'^-,J=0) & = & {1+P_1P_2\over2} {8\pi \over m_L^2} 
\left({\lambda_{ee}\lambda{\ell\ell'} \over \sum_{ij} \lambda_{ij}^2}\right)^2
\\
\label{xsv}
\sigma(\ee\to\ell^-\ell'^-,J=1) & = & {1-P_1P_2\over2} {48\pi \over m_L^2} 
\left({\lambda_{ee}\lambda{\ell\ell'} \over \sum_{ij} \lambda_{ij}^2}\right)^2
~,
\eea
where $\ell\ell'=e,\mu,\tau$.
The total number of events expected on any \dil\ peak
is thus of the order of
\beq
\label{events}
n = {\cal L} \sigma \approx 10^7 
{{\cal L}[\mbox{fb}^{-1}] \over m_L^2[\mbox{TeV$^2$}]}
\left({\lambda_{ee}\lambda{\ell\ell'} \over \sum_{ij} \lambda_{ij}^2}\right)^2
~,
\eeq
where the last factor is less or equal to one.
In any case,
whatever finite value it assumes,
a spectacular resonance is thus sure to be observed \cite{FN}.
Note also 
how conveniently a beam polarization flip
can discriminate
between scalar and vector \dil s.

If the leptonic couplings are so small,
that the \dil\ width drops far below the beam energy spread,
the signal will be attenuated accordingly.
Still,
the rates remain substantial,
even for minute leptonic couplings \cite{G}.

This reaction is so spectacular and unavoidable,
that we do not need to consider the production 
of doubly-charged \dil s 
in the other \lc\ modes.

It has also been advocated \cite{george}
to study the cross-channel flavour violating reaction 
$\mu^+e^- \to \mu^-e^+$,
if someday a muon and an electron collider can be combined.

\subsubsection{\ep\ Scattering}

\begin{figure}[h]
\unitlength.5mm
\SetScale{1.418}
\begin{boldmath}
\begin{center}
\begin{picture}(70,40)(0,0)
\ArrowLine(0,30)(15,15)
\Photon(0,0)(15,15){2}{4}
\ArrowLine(15,15)(45,15)
\ArrowLine(60,30)(45,15)
\DashLine(45,15)(60,0){1}
\Text(-2,0)[r]{$\gamma$}
\Text(-2,30)[r]{$e^-$}
\Text(62,0)[l]{$L^-$}
\Text(62,30)[l]{$\bar\nu$}
\end{picture}
\qquad\qquad
\begin{picture}(80,40)(0,0)
\ArrowLine(0,30)(30,30)
\ArrowLine(60,30)(30,30)
\Photon(0,0)(30,0){2}{5}
\DashLine(30,0)(60,0){1}
\DashLine(30,0)(30,30){1}
\Text(-2,0)[r]{$\gamma$}
\Text(-2,30)[r]{$e^-$}
\Text(62,0)[l]{$L^-$}
\Text(62,30)[l]{$\bar\nu$}
\end{picture}
\end{center}
\end{boldmath}
\caption{
  Lowest order Feynman diagrams 
  for the associate production 
  of a singly-charged \dil\ $L^-$
  and an anti-neutrino
  in \ep\ collisions.
  The produced \dil\
  can be either a scalar $L^-_1,L^-_3$ 
  or the vector $L^-_{2\mu}$.
  In the case of the $L^-_1$,
  the neutrino cannot be of the electron type.
}
\label{fep}
\end{figure}
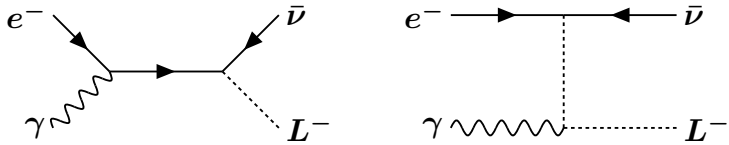

In principle,
doubly-charged \dil s can be produced in \ep\ collisions 
(\ref{depd})
with substantial \xs s \cite{LTNL}.
However,
these processes cannot compete
with the resonant production in \ee\ collisions
(\ref{deed}).
A singly-charged \dil,
though,
can also be produced in \ep\ scattering 
(\ref{deps})
as depicted in Fig.~\ref{fep}.
As these \dil s decay into a lepton and a neutrino,
the signal to be tagged
is a lepton and missing energy.
The analysis is more complex 
than in \ee\ scattering,
because 
\begin{itemize}
\item one has to fold the \xs s (\ref{eps}--\ref{epym})
over the photon energy and polarization spectra \cite{GIN};
\item there is no resonance and the signal is hence weak at threshold;
\item the \sm\ backgrounds from $W^-$ or $Z^0$ production 
and subsequent leptonic or invisible decays are substantial.
\end{itemize}

The polarized differential \xs s
for the production of scalar, vector or Yang-Mills fields
are given by

\bea
\label{eps}
&&
{d\sigma(\ep\to\bar\nu L^-) \over dt} 
=
{2\pi\alpha^2 \over s^2}
{\lambda^2 \over e^2}
{1-P_e \over 2}
{-u \over s(t-m_L^2)^2}
\\\nonumber&&\hskip2em
\left[
  {1-P_\gamma\over2} t^2 + {1+P_\gamma\over2} m_L^4
\right]
\eea

\bea
\label{epv}
&&
{d\sigma(\ep\to\bar\nu L^-_\mu,\kappa_\gamma=1) \over dt} 
=
{2\pi\alpha^2 \over s^2}
{\lambda^2 \over e^2}
{1+P_e \over 2}
{1 \over 4m_L^2s(t-m_L^2)^2}
\\\nonumber&&\hskip2em
\left\{
  {1-P_\gamma\over2} 
    u \left[ 8um_L^4 - tu(s-8m_L^2) - 2m_L^2(s-2m_L^2)^2 \right]
\right.
\\\nonumber&&\hskip2em
\left.
  - {1+P_\gamma\over2} 
    \left[ s^3t + 2m_L^2(s-2m_L^2)^2 \right]
\right\}
\eea

\bea
\label{epym}
&&
{d\sigma(\ep\to\bar\nu L^-_\mu,\kappa_\gamma=0) \over dt} 
=
{2\pi\alpha^2 \over s^2}
{\lambda^2 \over e^2}
{1+P_e \over 2}
{-2u \over s(t-m_L^2)^2}
\\\nonumber&&\hskip2em
\left[
  {1-P_\gamma\over2} (s-m_L^2)^2 + {1+P_\gamma\over2} (u-m_L^2)^2
\right]
~.
\eea

For completeness,
we also mention the integrated \xs s
for both the singly- and doubly-charged \dil s
and
for any value of the electric anomalous coupling $\kappa_\gamma$.
Defining
\beq
x = {m_L^2 \over s}
~,
\eeq
they are given by

\begin{eqnarray}
\makebox[0cm][l]{\hskip0mm$\displaystyle 
   \sigma(J=0) 
    = 
    {\pi\alpha^2\over2s} 
    {\lambda^2\over e^2}
    {1 \pm P_e \over 2}
    \times$}
\label{s0}\\
&& 
\biggl\{\quad
\left(-\left(3+4Q\right) + \left(7+8Q+8Q^2\right)x\right) 
\quad (1-x)
\nonumber\\
&&
+4Q\left( Q - \left(2+Q\right)x \right) 
\quad x \ln x
\nonumber\\
&&
-2\left(1+Q\right)^2\left( 1 - 2x + 2x^2 \right) 
\quad \ln{m_\ell^2/s\over\left(1-x\right)^2}
\nonumber\\
&&
\pm~P_\gamma~\Bigl[\quad
\left(-\left(7+12Q+4Q^2\right) + 3x\right) 
\quad (1-x)
\nonumber\\
&&
+4Q^2
\quad x \ln x
\nonumber\\
&&
-2\left(1+Q\right)^2\left( 1 - 2x \right) 
\quad \ln{m_\ell^2/s\over\left(1-x\right)^2}
\quad \Bigr] \quad \biggr\}
\nonumber\\
\nonumber\\
\makebox[0cm][l]{\hskip-0mm$\displaystyle 
   \sigma(J=1)
    = 
    {\pi\alpha^2\over8m^2} 
    {\lambda^2\over e^2}
    {1 \pm P_e \over 2}
    \times$}
\label{c0}\\
&&\hspace{-3em} 
\biggl\{\quad
\left(\left(32-32\kappa+\kappa^2\right)Q^2 + \left(8-16(2-\kappa)Q-\kappa^2Q^2\right)x + 8\left(7+8Q+8Q^2\right)x^2\right)
\quad (1-x) 
\nonumber\\
&&\hspace{-2em} 
-4Q\left( \kappa^3Q + \left(8(2-\kappa)+\kappa(4-\kappa)Q\right)x - 8Qx^2 + 8\left(2+Q\right)x^3 \right) 
\quad \ln x
\nonumber\\
&&\hspace{-2em} 
-16\left(1+Q\right)^2\left( 1 - 2x + 2x^2 \right) 
\quad x \ln{m_\ell^2/s\over\left(1-x\right)^2}
\nonumber\\
&&\hspace{-3em}  
\pm~P_\gamma~\Bigl[\quad
\left(- 3\kappa^2Q^2 + \left(40+16(2+\kappa)Q+(96-32\kappa-\kappa^2)Q^2\right)x + 24x^2\right)
\quad (1-x)
\nonumber\\
&&\hspace{-0em} 
-4Q\left( -(8+4\kappa+\kappa^2)Q + 8\left(4-\kappa+Q\right)x \right) 
\quad x \ln x
\nonumber\\
&&\hspace{-0em} 
+16\left(1+Q\right)^2\left( 1 - 2x \right) 
\quad x \ln{m_\ell^2/s\over\left(1-x\right)^2}
\quad \Bigr] \quad \biggr\}
\nonumber
~,
\end{eqnarray}
where $\kappa=\kappa_\gamma$
and $m_\ell=m_\mu,m_\tau$ is the mass of the $u$-channel charged lepton 
exchanged in the production of a doubly-charged \dil.
The unpolarized expressions
for the associated electron production
are given in Ref.~\cite{LTNL}.

The $Z^0$ background
is entirely confined in the region of phase space
where the energy $E_\ell$ and polar angle $\theta_\ell$
of the single emerging lepton
are contained within:
\beq
\label{kinz}
E_\ell
>
{y_{\rm min} s - m_Z^2 
\over
\sqrt{s} \left[ (1-\cos\theta_\ell) + y_{\rm min} (1+\cos\theta_\ell) \right]}
~,
\eeq
where $y=E_{\gamma}/E_e$ 
is the energy fraction
of the photons.
The exact value of its minimum can be tuned 
by changing the distance between the conversion and interaction points.
In the following 
we assume $y_{\rm min}=.5$.

Similarly,
most of the $W^-$ events
are located in the region of phase space where
\beq
\label{kinw}
E_\ell
<
{m_W^2 y \sqrt{s} 
\over
y^2 s (1+\cos\theta_\ell) + m_W^2 (1-\cos\theta_\ell)}
~.
\eeq

Since the vector \dil\ $L^-_{2\mu}$
couples to right-handed electrons,
polarizing accordingly the electron beam
will result into a further suppression of the $W^-$ background.
Unfortunately, 
the scalar \dil s $L_1^-$ and $L_3^-$
will need a left polarized electron beam
and will thus have to compete with the large $W^-$ background.

\begin{figure}[h]
\input{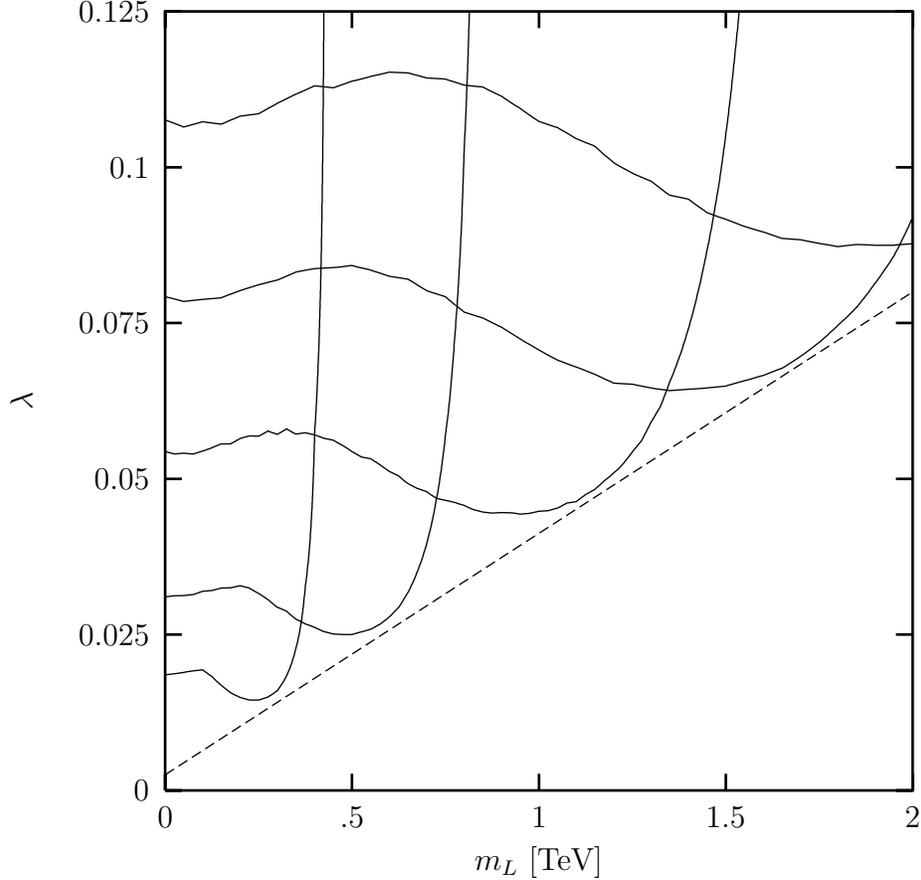}
\caption{
Smallest observable scalar \dil\ $L_1^{-}$ 
couplings to leptons
at the one standard deviation level.
as a function of the \dil\ mass
in \ep\ collisions.
The collider's \pe\ \cm\ energies
are .5, 1, 2, 3 and TeV
from left to right.
The other parameters used 
are given in Eqs~(\protect\ref{epas}).
}
\label{feg2nl}
\end{figure}

To gauge the discovery potential of \ep\ collisions,
we have plotted 
in Fig.\ref{feg2nl}
for the scalar \dil\ $L_1^{-}$
and several collider \cm\ energies
the $\chi^2_\infty=1$ boundary 
in the $(\lambda,m_L)$ plane
of the Cram\'er-Rao limit \cite{CR}
({\em cf.} Appendix B)
\beq
\label{creg}
\chi_\infty^2 = {\cal L} \int dy P(y) \int d\cos\theta_\ell
{
  \left( \displaystyle{d\sigma(\lambda) \over d\cos\theta_\ell} -
         \displaystyle{d\sigma(\lambda=0) \over d\cos\theta_\ell} \right)^2
  \over
  \displaystyle{d\sigma(\lambda=0) \over d\cos\theta_\ell}
}
~.
\eeq
The photon energy spectrum $P(y)$ is given in Ref.~\cite{GIN}
and the electron-photon \cm\ energy is 
$\sqrt{s_{e\gamma}} = \sqrt{ys_{ee}}$.
To obtain these results
we have included all decay channels of the \dil s
and have made the following realistic assumptions:
\bea
\label{epas}
&& {\cal L}_{e^+e^-} = 20 \mbox{ fb}^{-1} \\\nonumber
&& |P_e| = 90 \% \\\nonumber
&& |P_{\rm laser}| = 100 \% \\\nonumber
&& .5 \le y=E_\gamma/E_e \le .83 \\\nonumber
&& \theta_\ell \ge 5^o \\\nonumber
&& E_\ell \ge 10 \mbox{ GeV}
~,
\eea
where $\theta_\ell$ and $E_\ell$ are the polar angle and energy 
of the emerging lepton.

Similar plots are obtained for the vector \dil s.
In general,
these curves are closely osculated by the relation
\bea
\label{oscs}
\hspace{-1em}
\left( {\displaystyle\sum_\ell\lambda_1^{e\ell} \over m_L/\mbox{TeV}} \right)^2
=
\left( \sqrt{2}{\displaystyle\sum_\ell\lambda_3^{e\ell} \over m_L/\mbox{TeV}} \right)^2
&=& 
.15
~ {\chi_\infty^2 \over {\cal L}/{\rm fb}^{-1}}
\qquad
\left(
  m_L \le .6\sqrt{s_{ee}}
\right)
\\
\label{oscv}
\hspace{-1em}
\left( {\displaystyle\sum_\ell\lambda_2^{e\ell} \over m_L/\mbox{TeV}} \right)^2
&=& 
.03
~ {{\chi_\infty^2 \over {\cal L}/{\rm fb}^{-1}}}
\qquad
\left(
  m_L \le .3\sqrt{s_{ee}}
\right)
\ ,
\eea
where 
$\chi_\infty^2$ is the required number of standard deviations,
$\cal L$ is the integrated luminosity
and $\ell = e,\mu,\tau$.
These scaling relations
provide a convenient means to gauge 
the \dil\ discovery potential of \ep\ scattering.

Although the relations (\ref{oscs},\ref{oscv}) are only valid 
for \dil\ masses lighter than 60\%\ or 30\%\ of the collider energy,
heavier \dil s with stronger couplings
can also be probed up to the kinematical limit
$m_L \simeq .91 \sqrt{s_{ee}}$.

Single \dil\ production
can also take place 
via the same submechanism
with quasi-real electrons
in \pp\ collisions \cite{LTNL}
with quasi-real photons
in \pe\ \cite{LP',LTNL}
or $e^-p$ \cite{AP} collisions.
As expected,
the \xs s are accordingly smaller.
It is straightforward to translate the results of Ref.~\cite{AP}
for doubly-charged scalar \dil s ligther than 150 GeV,
into the approximate HERA exclusion range
$
\lambda \lsim 0.65 m_L
$
[TeV].

\begin{comment}{
\bea
\label{hera}
\left( {\displaystyle\sum_\ell\lambda_1^{e\ell} \over m_L/\mbox{TeV}} \right)^2
=
\left( \sqrt{2}{\displaystyle\sum_\ell\lambda_3^{e\ell} \over m_L/\mbox{TeV}} \right)^2
&\lsim& 
1.7
~.
\eea
}\end{comment}

\subsubsection{\pe\ Scattering}

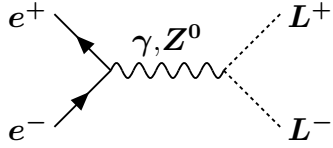
\begin{figure}[h]
\unitlength.5mm
\SetScale{1.418}
\begin{boldmath}
\begin{center}
\hskip1em
\begin{picture}(80,40)(0,0)
\ArrowLine(0,0)(15,15)
\ArrowLine(15,15)(0,30)
\Photon(15,15)(45,15){2}{5}
\DashLine(60,30)(45,15){1}
\DashLine(45,15)(60,0){1}
\Text(-2,0)[r]{$e^-$}
\Text(-2,30)[r]{$e^+$}
\Text(62,0)[l]{$L^-$}
\Text(62,30)[l]{$L^+$}
\Text(30,23)[c]{$\gamma$,${Z^0}$}
\end{picture}
\end{center}
\end{boldmath}
\caption{
  Lowest order Feynman diagrams contributing to $\pe\to L^+L^-$ scattering,
  where the final state \dil s
  can be either the scalars $L^{-}_1,L^{-}_3$ 
  or the vector $L^{--}_{2\mu}$.
}
\label{fpes}
\end{figure}

Doubly-charged \dil s can in principle be pair-produced in \pe\ scattering 
(\ref{dped})
\cite{S,GMS,GGMKO,LP'},
but these processes can under no circumstances compete
with the resonant production in \ee\ collisions
(\ref{deed}).
However,
the pair-production of singly-charged \dil s 
(\ref{dpes})
may become interesting
if their couplings to leptons turn out to be too small 
to be observed in \ep\ scattering
(\ref{deps}).
Indeed,
as depicted in the Feynman diagram of Fig.~\ref{fpes},
\dil s can still be produced 
thanks to their couplings to the neutral gauge bosons
(\ref{lagscal},\ref{lagvec}).

The signal will consist of 2-lepton events 
with missing energy.
The \sm\ backgrounds from $W$ pair production
are substantial,
but may be rendered harmless in \pe\ scattering
with right-handed electron beams.

Assuming the \dil s couple so weakly to leptons
that their discovery is precluded in \ep\ scattering,
we can ignore the $t$-channel lepton exchange in the \pe\ reaction.
Whatever the charge of the produced \dil s,
using the definitions (\ref{pol},\ref{sm})
the integrated scalar \cite{S,GMS,GGMKO,LP'} and vector 
\pe\ \xs s are then

\bea
\label{spe}
\hspace{-2em}
\sigma(\pe \to L \bar L)
&=&
{2\pi\alpha^2 \over 3s}
\beta^3
\Sigma
\\
\label{vpe}
\hspace{-2em}
\sigma(\pe \to L_\mu \bar L_\mu)
&=&
{\pi\alpha^2 \over 6s}
\beta^3
\Sigma
\left[ 4\kappa^2+4\kappa-1 
     - 3\beta^2
     + 4(\kappa-1)^2 {1 \over 1-\beta^2} \right]
~,
\eea
where
\beq
\label{equal}
\beta = \sqrt{1 - {4m_L^2 \over s}}
\qquad\qquad
\kappa = \kappa_\gamma = \kappa_Z
\eeq
and 
\beq
\Sigma
=
[LL] \left( \sum_{i=\gamma,Z^0} {s \over (s-m_i^2)} Q_i L_i \right)^2
+
[RR] \left( \sum_{i=\gamma,Z^0} {s \over (s-m_i^2)} Q_i R_i \right)^2
~.
\eeq

We do not know {\em a priori} the value 
of the electroweak anomalous coupling
$\kappa$
in Eq.~(\ref{vpe}).
The most conservative bounds 
to be expected from \pe\ scattering
on vector \dil s
are given by the value of $\kappa$
which minimizes the \xs\
\beq
\label{kkmin}
\kappa_{\rm min} = {1\over2} {1+\beta^2 \over 2-\beta^2}
~.
\eeq
It is only for simplicity,
that we assume the electric and weak anomalous coupling
to be equal in Eq.~(\ref{equal}).
It must be borne in mind
that some unfortunate combinations
may suppress the vector \xs s
even further.
We discard such a possibility here.

In the limits
\beq
\label{lim}
|P_{e^-}|=|P_{e^+}|=1
\qquad\qquad
m^2_Z \ll s
\qquad\qquad
\sin^2\theta_w={1\over4}
\eeq
the scalar and vector \xs s are then given by
\bea
\label{smin}
\sigma(e^+_Re^-_R \to L^+ L^-)
&=&
{32\pi\alpha^2 \over 27s}
\beta^3
\\\nonumber
\sigma(e^+_Le^-_L \to L^+ L^-)
&=&
{1\over4}
\sigma(e^+_Re^-_R \to L^+ L^-)
\\
\label{vmin}
\sigma_{\rm min}(e^+_Re^-_R  \to L^+_\mu L^-_\mu)
&=&
{2\pi\alpha^2 \over 3s}
\beta^3
{ 5 - 4\beta^2 + 3\beta^4 \over 2 - \beta^2}
\\\nonumber
\sigma(e^+_Le^-_L  \to L^+_\mu L^-_\mu) 
&=& 
0
~.
\eea
If the positrons cannot be polarized,
these \xs s are to be divided by 2.

\begin{figure}[h]
\input{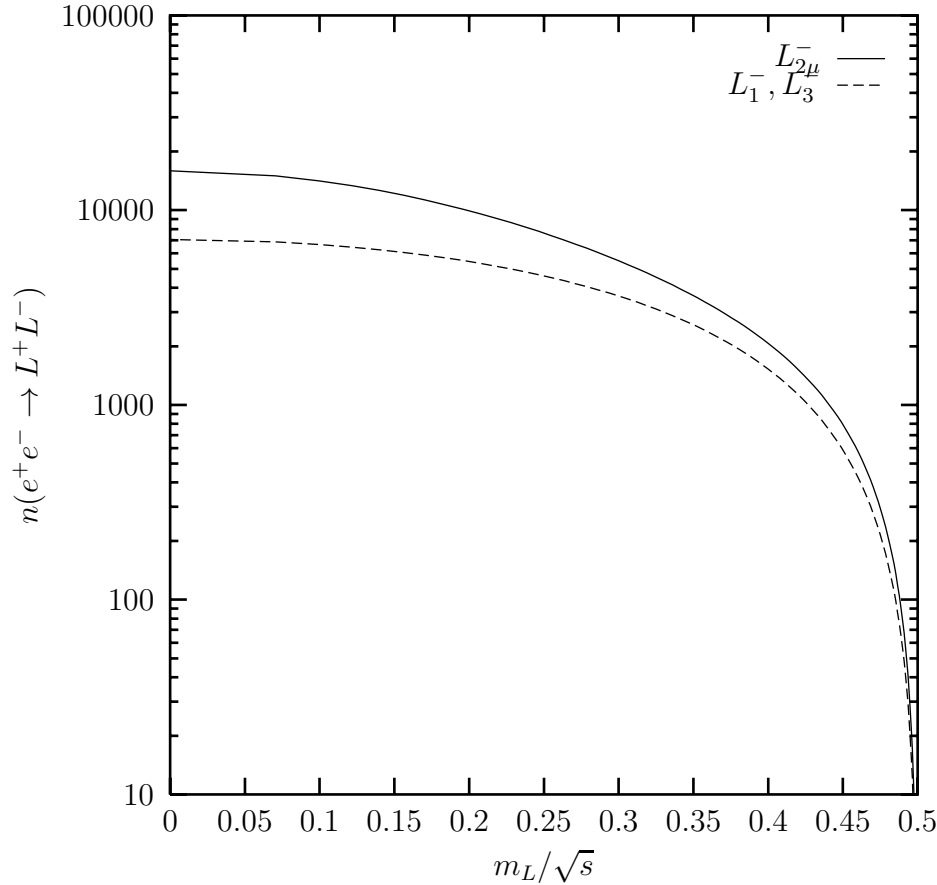}
\caption{
Mass dependence of the number of pair-produced
singly-charged \dil s in \pe\ annihilations,
assuming the luminosity scaling law
(\protect\ref{lumpe}).
}
\label{fpe2ll}
\end{figure}

Assuming the luminosity scaling law (\ref{lumpe})
and all \dil\ decay channels to be observed,
we plot in Fig.~\ref{fpe2ll}
the number of expected \dil\ events in right-polarized \pe\ collisions
as a function of the ratio of \dil\ mass to the \cm\ energy.
Clearly,
a significant signal is expected,
even close to the kinematical limit.

\subsubsection{\pp\ Scattering}

In principle \dil s can also be pair-produced in \pp\ collisions
(\ref{dped},\ref{dppd}).
However,
the doubly-charged \dil s will be much better observed in \ee\ scattering.
Similarly,
\ep\ collisions or
\pe\ annihilations with a right-handed electron beam
will offer a better signal to background ratio.
Moreover,
the $\gamma\gamma$ \cm\ energy 
cannot exceed {\em ca} 83\%\ 
of the \pe\ collider energy.
For all these reasons,
\pp\ collisions
are only of marginal interest
for discovering \dil s,
and we do not consider these reactions here.

\subsection{High Energy Summary}

The present and prospective high energy bounds on $L=2$ \dil s
are summarized in Tables~\ref{thei} and \ref{thed}.
Only the limits from the experiments which provide 
the best constraints on the \dil\ masses and couplings to leptons
are listed.
All bounds are stated at the 95\%\ confidence level.

\renewcommand{\arraystretch}{2}
\begin{table}
$
\setlength{\arraycolsep}{1em}
\begin{array}{||l||ccc|ccc||}
\hline
\hline
  & \multicolumn{3}{c|}{\ee}
  & \multicolumn{3}{c||}{\pe}
\\
  & \mbox{diagonal}
  & \mbox{democracy}
  & \mbox{infiltration}
  & \mbox{diagonal}
  & \mbox{democracy}
  & \mbox{infiltration}
\\
\hline
\hline
  &
  &
  & {\lambda^1 \over m}>{0.02 \over \sqrt{s}}
  &
  &
  & {\lambda^1 \over m}>{0.025 \over \sqrt{s}}
\\
  \tilde L_1^{--}
  & {\lambda   \over m}>{0.02 \over \sqrt{s}}
  & {\lambda   \over m}>{0.02 \over \sqrt{s}}
  & {\lambda^1 \over m}{\lambda^2 \over m}>{0.001\over {s}}
  & {\lambda   \over m}>{0.025\over \sqrt{s}}
  & {\lambda   \over m}>{0.025\over \sqrt{s}}
  & {\lambda^2 \over m}>{5    \over \sqrt{s}}
\\
  &
  &
  & {\lambda^1 \over m}{\lambda^3 \over m}>{0.001\over {s}}
  &
  &
  & {\lambda^3 \over m}>{82    \over \sqrt{s}}
\\
\hline
  &
  &
  & {\lambda^1 \over m}>{0.035\over \sqrt{s}}
  &
  &
  & {\lambda^1 \over m}>{0.025\over \sqrt{s}}
\\
  L_{2\mu}^{--}
  & {\lambda   \over m}>{0.035\over \sqrt{s}}
  & {\lambda   \over m}>{0.035\over \sqrt{s}}
  & {\lambda^1 \over m}{\lambda^2 \over m}>{0.003\over {s}}
  & {\lambda   \over m}>{0.025\over \sqrt{s}}
  & {\lambda   \over m}>{0.025\over \sqrt{s}}
  & {\lambda^2 \over m}>{8    \over \sqrt{s}}
\\
  &
  &
  & {\lambda^1 \over m}{\lambda^3 \over m}>{0.003\over {s}}
  &
  &
  & {\lambda^3 \over m}>{128  \over \sqrt{s}}
\\
\hline
  &
  &
  & {\lambda^1 \over m}>{0.015\over \sqrt{s}}
  &
  &
  & {\lambda^1 \over m}>{0.02 \over \sqrt{s}}
\\
  L_3^{--}
  & {\lambda   \over m}>{0.015\over \sqrt{s}}
  & {\lambda   \over m}>{0.015\over \sqrt{s}}
  & {\lambda^1 \over m}{\lambda^2 \over m}>{0.0007\over {s}}
  & {\lambda   \over m}>{0.02 \over \sqrt{s}}
  & {\lambda   \over m}>{0.02 \over \sqrt{s}}
  & {\lambda^2 \over m}>{3    \over \sqrt{s}}
\\
  &
  &
  & {\lambda^1 \over m}{\lambda^3 \over m}>{0.0007\over {s}}
  &
  &
  & {\lambda^3 \over m}>{58   \over \sqrt{s}}
\\
\hline
\hline
\end{array}
$
\vspace{9cm}
\caption{
Best explorable limits on the coupling to mass ratio
of the doubly-charged \dil s
in indirect high energy \ee\ and \pe\ searches.
These 95\%\ confidence bounds
are given for the three flavour coupling models 
(\protect\ref{diag}--\protect\ref{lap}).
We assume unpolarized positron beams 
and use the luminosities (\protect\ref{lumpe},\protect\ref{lumee}) 
for the \pe\ and \ee\ mode.
}
\label{thei}
\end{table}
\renewcommand{\arraystretch}{1}

\renewcommand{\arraystretch}{2}
\begin{table}
$
\setlength{\arraycolsep}{1em}
\begin{array}{||l||c|c|c|cc||}
\hline
\hline
  & \mbox{LEP }(Z^0)
  & \ee
  & \pe
  & \multicolumn{2}{c|}{\ep}
\\
\hline
\hline
  L_1^-
  & m<44 \mbox{ GeV}
  &
  & m<0.5\sqrt{s}
  & m<0.6\sqrt{s} 
  & {\sum\lambda \over m}>{0.04 \over \sqrt{s}}
\\
\hline
  \tilde L_1^{--}
  & m<45 \mbox{ GeV}
  & m<\sqrt{s}
  &&&
\\
\hline
  L_{2\mu}^-
  & m<45 \mbox{ GeV}
  &
  & m<0.5\sqrt{s}
  & m<0.3\sqrt{s} 
  & {\sum\lambda \over m}>{0.02 \over \sqrt{s}}
\\
\hline
  \tilde L_{2\mu}^{--}
  & m<38 \mbox{ GeV}
  & m<\sqrt{s}
  &&&
\\
\hline
  L_3^0
  & m<44 \mbox{ GeV}
  &&&&
\\
\hline
  L_3^-
  & m<44 \mbox{ GeV}
  &
  & m<0.5\sqrt{s}
  & m<0.6\sqrt{s} 
  & {\sum\lambda \over m}>{0.04 \over \sqrt{s}}
\\
\hline
  \tilde L_3^{--}
  & m<45 \mbox{ GeV}
  & m<\sqrt{s}
  &&&
\\
\hline
\hline
\end{array}
$
\vspace{1ex}
\caption{
Best explored or explorable \dil\ masses and couplings
in direct high energy searches.
These are at least 95\%\ confidence bounds.
The future collider limits 
assume the validity of the luminosity scaling law
(\protect\ref{lumpe}).
}
\label{thed}
\end{table}
\renewcommand{\arraystretch}{1}

The LEP experiments
constrain all \dil s to have a mass exceeding at least 44 GeV,
except for the doubly-charged vector $L_{2\mu}^{--}$
which couples very weakly to the $Z^0$ bosons
and could still be as light as 38 GeV.
These bounds are totally model-independent.

In high energy Bhabha and M\o ller scattering
virtual doubly-charged \dil s may induce 
corrections or even lepton flavour violations.
The non-observation
of such effects
at a \lc\ of the next generation
will allow stringent limits to be set on the ratio
of the leptonic coupling to the mass 
$\lambda / m_L$.
This will substantially improve the low-energy bounds.
Moreover,
the discovery prospects
are to a much lesser extent 
dependent on the structure of the flavour coupling matrix.

If the collider \cm\ energy reaches the mass of a doubly-charged \dil,
it will show up as a spectacular resonance in \ee\ scattering.
This is therefore 
the privileged mode for discovering and studying the properties 
of doubly-charged \dil s.

Singly-charged \dil s can be produced both in \pe\ and \ep\ scattering.
Provided the collider energy is sufficient
to pair-produce \dil s,
the \pe\ mode can probe \dil s which couple very weakly to leptons
and hence provide model-independent mass limits.
Heavier \dil s can be searched for in \ep\ reactions,
if their couplings to leptons are not too weak.

Once \dil s are produced on-shell,
the information gathered from 
their \xs s
and
their decay modes 
will unambiguously determine 
a large portion of the 
coupling constant matrix.

Unless a $Z'$ resonance is directly accessible,
standard high energy experiments
have no prospects 
for seeing the neutral \dil\ $L_3^0$.

\section{Conclusion}

We have derived the most general renormalizable
\dil\ lagrangians
consistent with the 
$SU(2)_L \otimes U(1)_Y$ symmetry.
Concentrating on those 7 \dil s
which carry lepton number $L=2$,
we have provided a compilation
of the present constraints 
from low-energy
experiments, and 
 the present and future bounds
which may be set by colliders. 

The model independent low-energy limits are listed in Table \ref{ler}.
They usually involve flavour non-diagonal couplings,
so their implications are difficult to judge. We make
three representative assumptions about the structure of
the coupling constant matrices, and present
the low-energy bounds with these assumptions in Tables
\ref{treq}, \ref{trdiag} and \ref{trafs}. Bileptons
with masses between 100 GeV and 10 TeV, and gauge
or Yukawa strength couplings would in general be
consistent with the data. 

Future high-energy \pe\ and \ee\ experiments 
at a linear collider of the next generation
may significantly extend
the present low-energy bounds,
as is summarized in Table~\ref{thei}.
Moreover,
if real \dil s can be produced on-shell
the observation of their decay modes 
will provide unambiguous information 
about the structure of the coupling constant matrix.
In particular,
the \ee\ \lc\ mode is ideally suited for searching 
doubly-charged \dil s, 
whereas singly-charged \dil s
can be well sought for in \pe\ or \ep\ collisions.

\section*{Acknowledgements}

We are very grateful to Milan Locher 
for his careful reading of the entire manuscript 
and his numerous critical comments and suggestions.
We also wish to thank David Bailey for useful conversations.

\section*{Appendices}

\begin{appendix}

\section{Four-Fermion Vertices}

Calculating bounds using
4-fermion vertices assumes that the
\dil\ masses are heavier than the energy scale of
the experiment, so that the momentum of
the \dil\ in the propagator can be neglected
\beq
\frac{1}{p^2 - m_L^2} \to \frac{ -1}{{m_L}^2}
\eeq

The four-fermion vertices listed
in the second column of Table~\ref{tff} 
are easily derived from the lagrangian (\ref{lag}).
They are of the form
\beq
\frac{a \lambda^2}{m_L^2} ( \bar{\psi} \Gamma \chi)(\bar{\epsilon}
\Gamma \delta) 
\label{star}
~,
\eeq
where $\psi, \chi, \epsilon$ and $\delta$ are {\em chiral}
fermions,  the $\Gamma$s are either
1 or $\gamma^{\mu}$, and $a$ includes
all the factors of 2.

The form (\ref{star})
may not convenient for comparing \dil\ rates with \sm\ ones.
However,
by appropriate Fierz transformations
and transpositions of matrix elements,
the relevant \dil\ four-fermion vertices
can all be brought in the \sm-like 
$(V \pm A)(V \pm A)$ or $(V + A)(V - A)$ forms.
For this operation
the following relations
turn out to be useful
\bea
\label{a1}
 (\bar{a}^c \gamma^{\mu} P_{L,R} b^c) 
  & = & - (\bar{b} \gamma^{\mu} P_{R,L} a) 
\\
\label{a2}
 (\bar{a} P_L b)(\bar{c} P_R d) & = & -\frac{1}{2}(\bar{a} \gamma^{\mu} P_R d)
   (\bar{c} \gamma_{\mu} P_L b) 
\\
\label{a3}
 (\bar{a} \gamma^{\mu} P_{L,R} b)(\bar{c} \gamma_{\mu} P_{L,R} d)
  & = & (\bar{a} \gamma^{\mu} P_{L,R} d)(\bar{c} \gamma_{\mu} P_{L,R}
b)
~.
\eea

Note that scalar and vector \dil s
do not induce any tensor matrix elements. The only
way to generate matrix elements of
the form $(\bar{\psi} \sigma^{\mu \nu} \chi)
(\bar{\epsilon} \sigma_{\mu \nu} \delta)$
 from the exchange of
a scalar or vector particle is
by Fierz-rearranging a scalar induced
four fermion vertex where the scalar
coupled to left-handed fermions at one end,
and right-handed at the other (operators
of the form $(\bar{\psi} R \chi)
(\bar{\epsilon} R \delta)$). 
The \sm\ symmetries  do not allow 
 scalar \dil s  with such interactions.

\section{The Cram\'er-Rao Limit}

The asymptotic resolution \cite{stats}
with which a reaction
can set bounds on a given parameter,
say the lepton-\dil\ coupling $\lambda$,
is given by the Cram\'er-Rao limit
\beq
\chi^2_\infty =
{\cal L}\
\int\!d\Omega\
{\left(
        \displaystyle{d\sigma(\lambda)\over d\Omega}
        -\displaystyle{d\sigma(0)\over d\Omega}
\right)^2
\over
        \displaystyle{d\sigma(0)\over d\Omega} 
}
~,\label{e7}
\eeq
where $\sigma(0)$ is the \sm\ expectation
and $d\Omega$ is a phase space element.
The $N$ standard deviation exclusion bounds for $\lambda$
are obtained by setting
$\chi^2_\infty = N^2$
in Eq.~(\ref{e7}).

If the systematic errors are small,
this limit is closely approached 
by a \ml\ estimator.
Indeed,
defining the probability density
\beq
p = {1\over\sigma} {d\sigma\over d\Omega}
~,\label{e81}
\eeq
for small values of $\lambda$
Eq.~(\ref{e7}) can be rewritten 
\beq
\chi^2_\infty 
=
n \lambda^2 \int\!d\Omega\ {1\over p}
\left.\left({\partial p \over \partial \lambda}\right)^2 \right|_{\lambda=0}
=
\lambda^2 \left<\left.-{\partial^2\ln L 
\over \partial \lambda^2}\right|_{\lambda=0}\right>
~,\label{e82}
\eeq
where $n$ is the total numebr of events.
Eq.~(\ref{e82}) defines the \ml\ estimator \cite{PDB},
where 
\beq
L = \prod_{i=1}^n\ p(\Omega_i)
\label{e83}
\eeq
is the \ml\ function.

To prove that this is indeed the Cram\'er-Rao minimum variance bound,
we set $\chi^2_\infty=1$ in Eq.~(\ref{e82}).
Discretizing into infinitesimal phase space bins 
labeled $i$,
we have
$p(\Omega_i)=n_i/n$
and we obtain for the inverse dispersion of $\lambda$
\beq
D(\lambda)^{-1}
=
\left.{1\over\lambda^2}\right|_{\chi^2_\infty=1}
=
\sum_i {1\over n_i} \left({\partial n_i \over \partial \lambda}\right)^2
~.\label{e91}
\eeq
By definition,
$n_i$ is the average number of events in bin $i$.
The observed number of events $N_i$ in this bin 
is distributed according to Poisson statistics,
{\em i.e.},
\beq
p_i = {n_i \over n} = {e^{-n_i} n_i^{N_i} \over N_i!}
\label{e92}
\eeq
is the probability to find $N_i$ events in bin $i$.
Assuming there are no bin-to-bin correlations,
we have
\bea
<N_i> \quad&=&\quad n_i
\label{e93}
\\
<(N_i-n_i)(N_j-n_j)> \quad&=&\quad \delta_{ij} n_i
\label{e94}
\eea
and we can rewrite Eq.~(\ref{e91})
\bea
D(\lambda)^{-1}
&=&
\sum_{i,j} \left< \left({N_i\over n_i}-1\right) \left({N_j\over n_j}-1\right) \right>
{\partial n_i\over\partial \lambda} {\partial n_j\over\partial \lambda} 
\nonumber\\
&=&
\left<\left( \sum_i \left({N_i\over n_i}-1\right) {\partial n_i\over\partial \lambda} \right)^2\right>
~.\label{e95}
\eea
This is nothing 
but the Cram\'er-Rao minimum variance bound
\beq
D(\lambda)^{-1}
=
\left<\left( \sum_i {\partial \ln p_i\over\partial \lambda} \right)^2\right>
~.\label{e96}
\eeq

To derive this result,
we only assumed the absence of bin-to-bin correlations 
in Eq.~(\ref{e94}).
Note that no assumption concerning the population of the bins is necessary.
Eq.~(\ref{e7}) provides thus a convenient means
for computing the Cram\'er-Rao bound of an experiment,
which in practice can be closely approached
by the \ml\ estimator
if the systematic errors are small.
In the presence of real data
the \ml\ function (\ref{e83}) can easily be evaluated
with all experimental resolutions and efficiencies folded in
\cite{tim}.

A more detailed treatment of this issue is provided in Ref.~\cite{CR}.

\end{appendix}

\end{document}